\documentclass[prl,twocolumn,superscriptaddress,floatfix, reprint]{revtex4-2}

\usepackage[usenames,dvipsnames]{color}
\usepackage[latin1]{inputenc}
\usepackage[english]{babel}
\usepackage{graphicx}
\usepackage{color}
\usepackage{amssymb,amsmath}
\usepackage[Gray,squaren]{SIunits}
\usepackage{xspace}
\usepackage{xcolor}
\usepackage{upgreek}
\usepackage{ulem}

\usepackage{amssymb,amsmath,verbatim,ulem}
\usepackage[export]{adjustbox}
\usepackage{dsfont}
\usepackage{appendix}
\usepackage{comment}
\usepackage[colorlinks = true,
           linkcolor = blue,
           urlcolor  = blue,
           citecolor = blue,
           anchorcolor = blue]{hyperref}
\newcommand{\ket}[1]{|{#1}\rangle}

\begin{document}

\title{Geometric Ramsey Interferometry with a Tripod Scheme}
\author{Chetan Sriram Madasu}
\email{chetansr001@e.ntu.edu.sg}
\affiliation{Nanyang Quantum Hub, School of Physical and Mathematical Sciences, Nanyang Technological University, 21 Nanyang Link, Singapore 637371, Singapore.}
\affiliation{MajuLab, International Joint Research Unit IRL 3654, CNRS, Universit\'e C\^ote d'Azur, Sorbonne Universit\'e, National University of Singapore, Nanyang Technological University, Singapore}
\affiliation{Center for Quantum Technologies, National University of Singapore, Singapore 117543, Singapore.}
\author{Ketan Damji Rathod}
\thanks{Currently at Bennett University, Greater Noida 201310, India.}
\affiliation{Center for Quantum Technologies, National University of Singapore, Singapore 117543, Singapore.}
\author{Chang Chi Kwong}
\affiliation{Nanyang Quantum Hub, School of Physical and Mathematical Sciences, Nanyang Technological University, 21 Nanyang Link, Singapore 637371, Singapore.}
\affiliation{MajuLab, International Joint Research Unit IRL 3654, CNRS, Universit\'e C\^ote d'Azur, Sorbonne Universit\'e, National University of Singapore, Nanyang Technological University, Singapore}
\author{David Wilkowski}
\affiliation{Nanyang Quantum Hub, School of Physical and Mathematical Sciences, Nanyang Technological University, 21 Nanyang Link, Singapore 637371, Singapore.}
\affiliation{MajuLab, International Joint Research Unit IRL 3654, CNRS, Universit\'e C\^ote d'Azur, Sorbonne Universit\'e, National University of Singapore, Nanyang Technological University, Singapore}
\affiliation{Center for Quantum Technologies, National University of Singapore, Singapore 117543, Singapore.}

\date{\today}
\begin{abstract}
Ramsey interferometry is a key technique for precision spectroscopy and to probe the coherence of quantum systems. Typically, an interferometer is constructed using two quantum states and involves a time-dependent interaction with two short resonant electromagnetic pulses. Here, we explore a different type of Ramsey interferometer where we perform quantum state manipulations by geometrical means, eliminating the temporal dependence of the interaction. We use a resonant tripod scheme in ultracold strontium atoms where the interferometric operation is restricted to a two-dimensional dark-state subspace in the dressed-state picture. The observed interferometric phase accumulation is due to an effective geometric scalar term in the dark-state subspace, which remarkably does not vanish during the free evolution time when the light-matter interaction is turned off. This study opens the door for more robust interferometers operating on multiple input-output ports. 
\end{abstract}

\maketitle

Ramsey interferometry employs temporally separated electromagnetic pulses to probe the energy difference and the coherence between two quantum states \cite{Ramsey1949ResonanceMethod, Ramsey1985MolecularBeams}.
Whether employing internal, external or both states of atoms, Ramsey interferometers become essential tools to probe quantum states in quantum simulations \cite{Cetina2016ManyBodyInterferometry, Li2016WilsonLine}, in quantum computing \cite{Lee2005PhaseControlOfTrappedIonQG}, in interband spectroscopy \cite{Dong2018RIofMotionalQuantumStates} and in atomic clocks at or below the quantum projection noise limit \cite{Santerelli1999QPN, Pedrozo2020EntanglementOnClockTransition}, to name a few.

In contrast to the majority of Ramsey interferometers that rely on the dynamical evolution of the system mediated through light-matter interaction, we explore here a geometric Ramsey interferometer governed by adiabatic evolution in the degenerate dark-state subspace of a tripod scheme. We find that the phase accumulation during the free evolution time arises from a geometric scalar potential. Surprisingly, this potential retains its physical significance even when the pulses are turned off as long as the dressed states of interest remain adiabatically connected to the bare states. Geometric scalar potentials are at the origin of the so-called dark optical lattices \cite{Dum1996GaugeStructuresInALI, Dutta1999ScalarTerm, Anderson2020SubwavelengthOpticalLattice}, and have been employed to create subwavelength barriers in an effective spin \cite{10.21468/SciPostPhys.11.6.100} or spinless \cite{Wang2018DarkStateOpticalLattice} configuration. Though the geometric scalar potential plays a role in shaping periodic potential, it is essentially overlooked in the bulk because of the moderate strength in comparison with commonly used optical potentials \cite{dalibard2011colloquium, Toyoda2013HolonomicSingleQubitOperations}.


\begin{figure}
    \centering
    \includegraphics[scale=1]{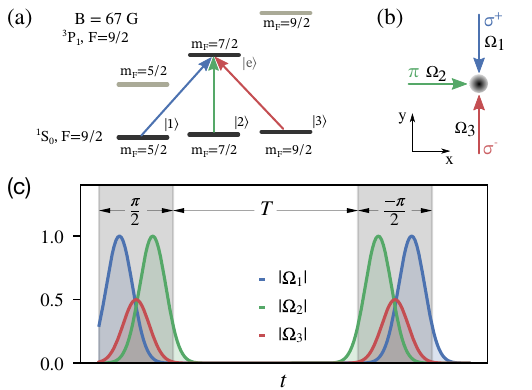}
    \caption{Schematic showing the implementation of the geometric Ramsey interferometry pulse sequence. (a) Energy levels of $^{87}\mathrm{Sr}$ atoms involved in the tripod scheme. A bias magnetic field of $67\,$G shifts adjacent excited magnetic states by approximately $930\Gamma$, where $\Gamma/2\pi=7.5\,$kHz is the linewidth of the intercombination line. (b) The spatial configuration of the tripod beams. (c) Relative Rabi frequencies of the tripod beams as a function of time. The Gaussian pulses are parameterized as $\Omega_a(t) = |\Omega_{0a}|e^{-(t-t^{(j)}_a)^2/4\sigma_t^2}$ where $|\Omega_{0a}|$ is the peak Rabi frequency with $a=1,2,3$, $\sqrt{2}\sigma_t$ is the temporal standard deviation and $t^{(j)}_a$ are the centers of the Gaussian pulses for the $\pi/2$ pulse ($j=1$) and $-\pi/2$ pulse ($j=2$). The pulse sequence corresponds to $t^{(j)}_1 = t^{(j)}_3 - \eta\sigma_t$, $t^{(j)}_2 = t^{(j)}_3 + \eta\sigma_t$, $t^{(1)}_3=4\sigma_t$ and $t^{(2)}_3=t^{(1)}_3+8\sigma_t+T$, with $|\Omega_{01}|=|\Omega_{02}|= 2|\Omega_{03}|\approx 2\pi\times260\,$kHz and $\eta$ is the separation parameter with a value of 1.8. Here, the length of $\pi/2$ pulses is defined as duration of the $\sigma^-$ Gaussian pulse i.e., $8\sigma_t$. Therefore, the separation between two $\pi/2$ pulses T, is defined as the free evolution time.}
    \label{Figure1}
\end{figure}


The interferometer operates on an ultracold gas of $^{87}\textrm{Sr}$ atoms. The gas is prepared using a two-stage magneto-optical trap \cite{chaneliere2008three, Tao2015HighFlux}, followed by evaporative cooling in a crossed-beam optical-dipole trap \cite{hasan2022anisotropic}. We then obtain a quantum degenerate Fermi gas comprised of $N=4.5(2)\times10^4$ atoms in the $m_F=9/2$ stretched Zeeman substate at a temperature of $T_0= 50(3)\,$nK. This temperature corresponds to $T_0/T_F = 0.25(2)$ where $T_F$ is the Fermi temperature.  Additionally, $T_0/T_R= 0.21(2)$, where $T_R$ is the recoil temperature associated with the tripod transitions. After evaporative cooling, the optical trap is switched off, and a magnetic field bias is turned on to isolate a tripod scheme on the $^1S_0, F_g=9/2 \to\,^3 P_1, F_e=9/2$ hyperfine multiplet of the intercombination line at $689\,$nm \cite{Hasan_2022}. Three laser beams resonantly couple the three internal ground states $|a \rangle \equiv |F_g, m_F\rangle$, with $a=\{1,2,3\}$ and $m_F=\{5/2, 7/2, 9/2\}$, respectively, to a common excited state $\ket{e}\equiv\ket{F_e, m_F=7/2}$, as shown in Figs. \ref{Figure1}a$\&$b. The light-matter interaction is characterized by three complex Rabi frequencies $\Omega_a$, associated with the $\ket{a}\to\ket{e}$ transitions.

\begin{figure}
    \centering
    \includegraphics[scale=1]{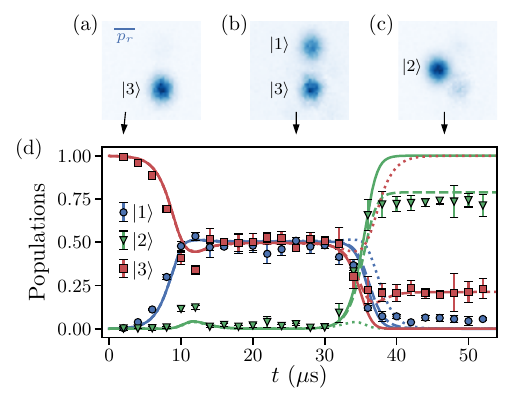}
    \caption{(a-c) Fluorescence images of the ultracold gas after $9\,$ms of time of flight, see \cite{SM} for more details. The images are taken before the first pulse ($t=0$), during the free evolution time ($t=18\,\mu$s), and after the second pulse ($t=48\,\mu$s), respectively. Each peak in the momentum distribution is associated with a bare state as indicated in each panel. We extract the bare state populations by fitting each peak to a 2D-Gaussian distribution.  (d) Populations of the bare states during the interferometric sequence, with $\sigma_t=2.5\,\mu$s and $T=6\,\mu$s. The experimental data points are plotted with markers with the error bars representing one standard deviation confidence. The plain and dashed curves represent the numerical integration of Eq. (\ref{unitaryT}) for temperatures of 0 and 50 nK, respectively. The dotted curve represents the zero temperature theoretical expectations, without the scalar term $\hat{Q}$.}
    \label{Figure2}
\end{figure}

Our geometric Ramsey interferometric sequence consists of a $\pi/2$ pulse and a $-\pi/2$ pulse, temporally separated by a free evolution time $T$ as sketched in Fig. \ref{Figure1}c. The first $\pi/2$ pulse, composed of three Gaussian pulses, puts the atoms initially in the $\ket{3}$ state into a coherent superposition of $\ket{3}$ and $\ket{1}$ states, ideally with equal probabilities. The relative population of the output states does not depend on the pulse duration, due to its geometrical nature. It is instead controlled by the relative peak Rabi frequency amplitude $|\Omega_{03}|$ of beam $3$ with respect to the peak Rabi frequency amplitudes of beams $1$ and $2$, which are set equal, namely $|\Omega_{01}|=|\Omega_{02}|$ \cite{Chetan2022AtomtronicDDT}. The second pulse, closing the interferometer, is a $-\pi/2$ pulse, meaning that, without any further phase accumulation, the second pulse brings back the atom into its initial state, namely $\ket{3}$. Here, an extra phase accumulation between the two arms occurs reducing the population of $\ket{3}$ at the interferometer output, as shown in Fig. \ref{Figure2} (red squares). Importantly, we note that the remaining population, instead of going to the state $\ket{1}$ (blue circles) as expected for a standard two-level Ramsey interferometer, is now transferred to state $\ket{2}$ (green triangles). This unusual behavior originates from the order of the Gaussian pulses acting on the tripod scheme. As shown in Fig. \ref{Figure1}c, the second pulse sequence is a temporal mirror image of the first one, so the beam $1$ pulse, which is finishing the sequence, prevents population in state $\ket{1}$ as expected for any STIRAP scheme \cite{Vitanov2017STIRAPReview}.

To confirm the phase-sensitive nature of the experiment, we purposely apply a phase jump $\Phi$ to the beam $3$ at the center of the free evolution sequence when the laser is considered to be turned off. As expected, the interferometric readout from the atomic bare state populations after the second pulse shows a sinusoidal evolution as a function of the introduced phase jump (see Fig. \ref{Figure3}a). We note that a phase jump of $\Phi=\pi$ rotates the interferometer output fringe by half a period.

During the free evolution time $T$, a phase accumulation occurs, which has a simple physical origin in the bare-state picture. The coherent transfer between the state $\ket{3}$ and the state $\ket{1}$ redistributes a photon between the beams $3$ and $1$, which leads to a momentum kick of $2p_r\hat{y}$ on the atom, as shown in Fig. \ref{Figure2}b. $p_r=\hbar k$ is the momentum recoil associated with a $\lambda=689\,$-nm photon, $\hbar$ is the reduced Planck constant, and $k=2\pi/\lambda$ is the wave number of the light field. Therefore, the phase accumulation corresponds to $\Delta E_k T/\hbar$ where $\Delta E_k=2p_r^2/m$ is the kinetic energy difference between the two bare states and $m$ is the atomic mass. We fit the Ramsey interferometer output evolution as a function of $T$ with a damped oscillation and find a frequency of $2\pi\times19.8(16)\,$kHz [see black solid curves in Fig. \ref{Figure3}b], in agreement with the theoretical prediction of $\Delta E_k/\hbar=2\pi\times 19.2\,$kHz.

At finite temperature, the oscillation is damped due to the momentum dispersion of the gas. In Fig. \ref{Figure3}b, we observe a good agreement of the experiment with an adiabatic model at a temperature $T_0=50\,$nK (see colored dashed curves), indicating that the temperature is the dominant dephasing mechanism limiting the experimental coherence time. Deviations from the model prediction are mainly due to a residual diabatic contribution; for more details, see \cite{SM}. The damping time, extracted from the fit, is found to be $\tau=23(4)\,\mu$s. Since $\tau\Delta E_k/\hbar\gtrsim 1$, only a few oscillations are visible, limiting the sensitivity of the frequency measurement. The coherence time can be improved either by reducing the temperature using for example delta-kick cooling \cite{PhysRevLett.78.2088} or by post-selection of a narrow momentum window after a long time of flight \cite{valdes-curiel_topological_2021}. Alternatively, Mach-Zehnder or Ramsey-Bord\'e types of interferometric pulse sequence can be in principle implemented to limit the dephasing due to temperature.


A rigorous theoretical treatment of the geometric Ramsey interferometer can be done with a brute-force diagonalization of the time-dependent Hamiltonian of the system. However, physical interpretation together with significant simplifications are possible by changing the original bare-state basis to the dressed-state basis of the internal Hamiltonian, defined by two long-lived zero-energy dark states, namely
\begin{eqnarray}
    \label{DarkStateBasis}
	\ket{D_1(\textbf{r},t)} &=&\sin \varphi(t) e^{2iky} \ket{1}-\cos\varphi(t) e^{ik(y-x)} \ket{2} \nonumber\\
	\ket{D_2(\textbf{r},t)} &=&\cos \vartheta(t)(\cos\varphi(t) e^{2iky}\ket{1} + \sin\varphi(t) e^{ik(y-x)} \ket{2})\nonumber\\ 
 &-& \sin \vartheta(t) \ket{3},
\end{eqnarray}
and two bright states that contain the bare excited state, so subject to a fast decay by photon spontaneous emission. Moreover, the bright states are light shifted by $\pm\hbar\Omega$, where $\vartheta = \cos^{-1}\left(|\Omega_3|/\Omega\right)$, $\varphi=\tan^{-1}\left(|\Omega_2|/|\Omega_1|\right)$, and $\Omega=\sqrt{|\Omega_1|^2+|\Omega_2|^2+|\Omega_3|^2}$ \cite{dalibard2011colloquium}.

\begin{figure}
    \centering
    \includegraphics[scale=1]{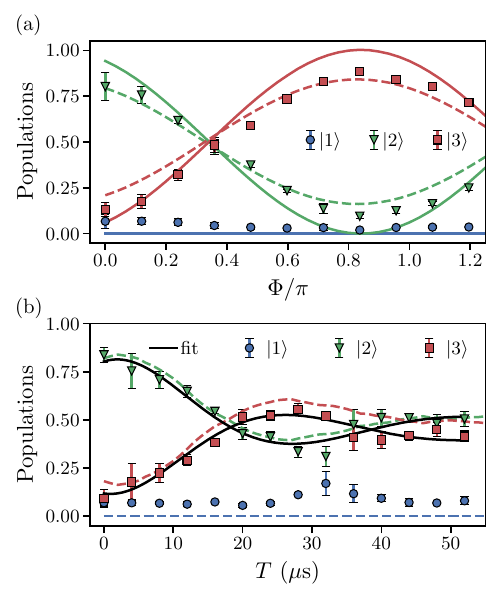}
    \caption{(a) Interference fringes generated with an abrupt phase change of the beam $3$ coupling the $\ket{3}\to\ket{e}$ transition. This phase jump $\Phi$ is introduced at the middle of the free evolution time, $T = 6\,\mu$s. The solid and dashed curves represent the numerical integration for temperatures of 0 and 50 nK, respectively. (b) Populations of bare states after the Ramsey pulse sequence as a function of free evolution time $T$. The black solid curves represent a fit using an exponentially damped oscillation. The colored dashed curves are the predictions of our adiabatic model; see Ref. \cite{SM} for more details.}
    \label{Figure3}
\end{figure}

A first simplification occurs because the internal state evolution can be restricted to the dark-state subspace. Indeed, the bright-state light shift corresponds to the highest energy scale of the problem ($\Omega\simeq 2\pi\times 410\,$kHz), and the initial bare state $\ket{3}$ is adiabatically connected to $\ket{D_2}$ \cite{Frederic2018Non-Abelian}. Overall, the populations of the bright states remain negligible during the Ramsey sequence. This point is experimentally checked noticing that there is no significant heating of the gas after the Ramsey sequence (for more details see Ref. \cite{SM}). Limiting ourselves now to the dark-state subspace, the effective Hamiltonian reads \cite{Ruseckas2005Non-Abelian, Frederic2018Non-Abelian}
\begin{equation}
    \label{Hamiltonian}
    \hat{H}=\frac{\hat{\textbf{p}}^2\otimes\mathds{1}}{2m}-\frac{\hat{\textbf{A}}\cdot\hat{\textbf{p}}}{m}+\hat{Q}+\hat{w},
\end{equation}
where $\mathds{1}$ is a two-dimensional identity operator defined in the dark-state subspace, and the operators $\hat{\textbf{A}}$, $\hat{Q}$, and $\hat{w}$ have the respective matrix entries
\begin{eqnarray}
    \label{AQw}
	\textbf{A}_{\mu\nu} &=&i\hbar\langle D_\mu|\boldsymbol{\nabla}D_\nu\rangle \nonumber \\
	Q_{\mu\nu} &=& \frac{\hbar^2}{2m}\langle \boldsymbol{\nabla}D_\mu|\boldsymbol{\nabla}D_\nu\rangle\nonumber\\ 
	w_{\mu\nu} &=& -i\hbar\langle D_\mu|\frac{\partial}{\partial t}D_\nu\rangle.
\end{eqnarray}
Since $|\boldsymbol{\nabla}|\sim k$ and the size of the momentum distribution is smaller than the recoil momentum $p_r$, as a second simplification, we neglect the kinetic and spin-orbit coupling contributions with respect to the scalar term $\hat{Q}$, i.e., the first and second right-hand-side terms of the Hamiltonian in Eq. (\ref{Hamiltonian}), respectively. The state evolution in the dark-state subspace is then given by the unitary transformation
\begin{equation}
	\hat{U}(t)=\mathcal{T}\exp{\left[-i\int_0^{t}\left(\hat{Q}(t^\prime)+\hat{w}(t^\prime)\right)\textrm{d}t^\prime\right]}.
	\label{unitaryT}
\end{equation}
where, $\mathcal{T}$ is the time-ordering operator. From the spatial configuration of our tripod beams (see Fig. \ref{Figure1}b), we derive the following expression for the scalar term
\begin{equation}
	\hat{Q} = -\frac{p_r^2}{2m}\left(\begin{array}{cc}
				2(1+\sin^2\varphi) & \cos\vartheta \sin 2\varphi \\
				\cos\vartheta \sin 2\varphi & 2\cos^2\vartheta(1+\cos^2\varphi)
		\end{array}\right)
    \label{GeometricScalarTerm}
\end{equation} 
and the final term on the right-hand side of Eq. (\ref{Hamiltonian}) reads
\begin{equation}
    \label{w_operator}
    \hat{w}=\hbar\cos \vartheta\frac{\partial \varphi}{\partial t}\hat{\sigma}_y,
\end{equation}
where $\hat{\sigma}_y$ is the $y$-component Pauli matrix. The operator $\hat{w}$ plays a key role since it is responsible for the geometric atomic beam splitting \cite{Vaishnav2008DDT,Toyoda2013HolonomicSingleQubitOperations,Chetan2022AtomtronicDDT}. We also note that this term has no specific energy scale since it depends on the temporal profile of the Gaussian pulse. The latter has to be slow enough to fulfill the adiabatic condition, namely $\langle \hat{w}\rangle\ll\hbar\Omega$, at all times. 

The solid curves in Fig. \ref{Figure2}d and Fig. \ref{Figure3} are obtained through numerical integrations of Eq. (\ref{unitaryT}), whereas the projections onto the bare states are extracted from Eq. (\ref{DarkStateBasis}). The dashed curves are obtained by averaging over the momentum distribution of our thermal sample, with a temperature of $T_0=50\,$nK. Here the momentum dependence is obtained in the semiclassical limit by reintroducing the previously overlooked first and second terms on the right-hand side of Eq. (\ref{Hamiltonian}); see main text and Ref. \cite{SM} for more details.

Our model, together with the damping due to the finite temperature, captures the main experimental features well, opening the door for insightful physical interpretations of this geometric Ramsey interferometer. As we have already mentioned, the initial state $\ket{3}$ is connected to $\ket{D_2}$ dark state \cite{Frederic2018Non-Abelian}. For a complete description, we shall highlight that the dark states at the end of the $\pi/2$ pulse are asymptotically connected to the bare states as $\ket{D_1}\rightarrow\ket{1}$ and $\ket{D_2}\rightarrow\ket{3}$ \cite{Chetan2022AtomtronicDDT}. This point can be easily verified, using Eq. (\ref{DarkStateBasis}) and noticing that at the end of the $\pi/2$ pulse $\varphi\rightarrow \pi/2$ and $\vartheta\rightarrow \pi/2$. Similarly, at the end of the $-\pi/2$ pulse, 
the dark states are connected to the bare states as $\ket{D_1}\rightarrow\ket{2}$ and $\ket{D_2}\rightarrow\ket{3}$. Hence, we understand that even if the geometric Ramsey interferometer is fundamentally a two-level interferometer in the dark-state subspace, we still need the three bare-ground states for a complete description. It leads to a multiple input-output port device, where the matter-wave propagation direction can be controlled by the pulse ordering sequence and a phase-sensitive signal (compare the bare-state population distribution locations in Fig. \ref{Figure2}a-c). This principle can be utilized for implementing an atomtronic bilateral switch where either the phase jump $\Phi$ or the free evolution time $T$ can be used as the switching control parameter.


Another insightful interpretation of our model concerns the nature of the phase accumulation during the free evolution time, which can be clearly associated with the scalar term $\hat{Q}$. Indeed, during the free evolution time $\partial\varphi/\partial t\rightarrow 0$, so $w_{\mu\nu}\rightarrow 0$. The last remaining term, which is the scalar potential, takes the asymptotic expression 
\begin{equation}
	 \lim_{\varphi\to \pi/2, \vartheta\to \pi/2} \hat{Q}=-\frac{p_r^2}{m}\left(\begin{array}{cc}
				2 & 0 \\
				0 & 0
		\end{array}\right).
    \label{GeometricScalarTerm_0}
\end{equation} 
Moreover, the dotted curves in Fig. \ref{Figure2}d correspond to numerical integrations of Eq. (\ref{unitaryT}) setting $\hat{Q}=\hat{0}$ at all times. Here, no phase shift is observed as the population transfers back to $\ket{3}$ at the output of the interferometer. A similar situation occurs with trapped ions in the Lamb-Dicke regime \cite{Toyoda2013HolonomicSingleQubitOperations}. We note that the presence of a nonzero $\hat{Q}$ term leads to a nonintuitive situation where the dressed-state picture remains meaningful even if the tripod beams are turned off, provided that the adiabatic asymptotic connection, depicted by Eq. (\ref{GeometricScalarTerm_0}), is fulfilled. In addition, the energy difference between the states $\ket{D_1}$ and $\ket{D_2}$ leads to a phase accumulation during the free evolution time of $|\hat{Q}_{11}-\hat{Q}_{22}|T/\hbar=2p_r^2T/\hbar m $ in agreement with the previously discussed bare-state approach.

\begin{figure}
    \includegraphics[scale=1]{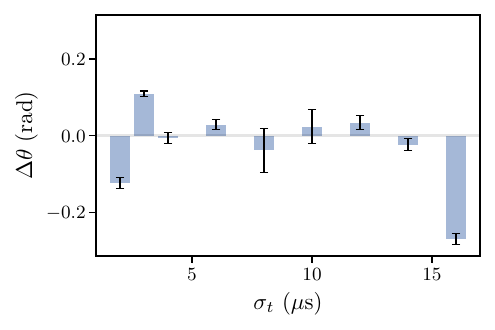}
    \caption{Deviation of the polar angle of the dark-state coherent superposition after the first $\pi/2$ pulse as a function of $\sigma_t$. }
    \label{Figure4}
\end{figure}

Finally, we check the geometrical nature of the matter-wave splitter, searching for time-independent behavior by either compressing or inflating the temporal sequence of the matter-wave splitter. For instance, the first $\pi/2$ pulse will rotate the initial $\ket{D_2}\equiv\ket{3}$  by a polar angle $\theta=\pi/2$ into the equatorial plane in the Bloch sphere representation. We show in Fig. \ref{Figure4} the deviation  $\Delta \theta = \theta_{exp}-\pi/2$ of the experimentally measured polar angle $\theta_{exp}$,  as a function of the temporal standard deviation of the Gaussian pulses $\sigma_t$. When the duration of the pulse sequence is within $3\ \mu \mathrm{s}<\sigma_t <15\ \mu \mathrm{s}$, the deviation is in agreement with a null value, indicating a time-independent geometric matter-wave splitter. For $\sigma_t <3\ \mu \mathrm{s}$, the  nonzero deviation indicates that the pulse sequence is not fully adiabatic. For $\sigma_t >15\ \mu \mathrm{s}$, the adiabatic approximation is fulfilled, but $\hat{w}$ becomes too small with respect to the spin-orbit and kinetic terms of Eq. (\ref{Hamiltonian}), leading to a breakdown of the approximation of our model given by Eq. (\ref{unitaryT}) \cite{Chetan2022AtomtronicDDT}. We use $\sigma_t=2.5\ \mu$s in the experiment in order to reduce the length of the pulse sequence as a trade-off between nonadiabaticity and effects of thermal dispersion. Under this condition, we check that the spontaneous emission is weak, which is a good indication that the adiabatic approximation remains correct; for more details see Ref. \cite{SM}.

In conclusion, we have explored a geometric Ramsey interferometer based on a tripod scheme. This interferometer reduces to a two-level system in the dark-state subspace but can also be viewed as connecting the three internal ground-bare states in a configuration with multiple input-output ports. We show that the phase accumulation during the free-evolution time is due to a geometric scalar potential that encapsulates the kinetic energy difference of the bare states. 
Because these states are time independent, geometric manipulations of quantum states are generally more robust than their dynamical counterparts. This robustness can be translated here to an interferometer that is insensitive to the mean velocity of the atomic ensemble, making it suitable for possible applications in quantum simulations and computing, and atomtronics circuits \cite{Wang2018DarkStateOpticalLattice, Kubala2021OpticalLatticeForTripod, Tsui2020StroboscopicOL,Frazer20221DArrays,amico2021roadmap}.

In the future, other types of interferometers, such as Ramsey-Bord\'e interferometers \cite{borde1984optical,borde1989atomic} or Mach-Zehnder interferometers \cite{PhysRevLett.67.181} can be envisioned using similar geometric approaches. The former can be utilized for precision measurements of the photon recoil shift to determine the fine-structure constant \cite{parker2018measurement,morel2020determination}, while the latter can serve for inertial sensing applications such as gravimetry \cite{menoret2018gravity}, gradiometry \cite{rosi2014precision}, or tests of the equivalence principle \cite{asenbaum2020atom}, to name a few. Finally, the inherent slow response time of adiabatic transformation can be addressed using shortcuts to adiabaticity schemes \cite{PhysRevA.95.043608}, enabling the implementation of large-area interferometers \cite{PhysRevLett.115.103001}. 

The authors thank Du Jinyi and Lucas Gabardos for careful reading of the manuscript. This work was supported by the CQT/MoE (Grant No. R-710-002-016-271), the Singapore Ministry of Education Academic Research Fund Tier2 (Grant No. MOE-T2EP50220-0008), and the Temasek Laboratories (Grant No. TLSP23-08).
\bibliography{GeometricRamsey}

\begin{thebibliography}{43}%
\makeatletter
\providecommand \@ifxundefined [1]{%
 \@ifx{#1\undefined}
}%
\providecommand \@ifnum [1]{%
 \ifnum #1\expandafter \@firstoftwo
 \else \expandafter \@secondoftwo
 \fi
}%
\providecommand \@ifx [1]{%
 \ifx #1\expandafter \@firstoftwo
 \else \expandafter \@secondoftwo
 \fi
}%
\providecommand \natexlab [1]{#1}%
\providecommand \enquote  [1]{``#1''}%
\providecommand \bibnamefont  [1]{#1}%
\providecommand \bibfnamefont [1]{#1}%
\providecommand \citenamefont [1]{#1}%
\providecommand \href@noop [0]{\@secondoftwo}%
\providecommand \href [0]{\begingroup \@sanitize@url \@href}%
\providecommand \@href[1]{\@@startlink{#1}\@@href}%
\providecommand \@@href[1]{\endgroup#1\@@endlink}%
\providecommand \@sanitize@url [0]{\catcode `\\12\catcode `\$12\catcode
  `\&12\catcode `\#12\catcode `\^12\catcode `\_12\catcode `\%12\relax}%
\providecommand \@@startlink[1]{}%
\providecommand \@@endlink[0]{}%
\providecommand \url  [0]{\begingroup\@sanitize@url \@url }%
\providecommand \@url [1]{\endgroup\@href {#1}{\urlprefix }}%
\providecommand \urlprefix  [0]{URL }%
\providecommand \Eprint [0]{\href }%
\providecommand \doibase [0]{https://doi.org/}%
\providecommand \selectlanguage [0]{\@gobble}%
\providecommand \bibinfo  [0]{\@secondoftwo}%
\providecommand \bibfield  [0]{\@secondoftwo}%
\providecommand \translation [1]{[#1]}%
\providecommand \BibitemOpen [0]{}%
\providecommand \bibitemStop [0]{}%
\providecommand \bibitemNoStop [0]{.\EOS\space}%
\providecommand \EOS [0]{\spacefactor3000\relax}%
\providecommand \BibitemShut  [1]{\csname bibitem#1\endcsname}%
\let\auto@bib@innerbib\@empty
\bibitem [{\citenamefont {Ramsey}(1949)}]{Ramsey1949ResonanceMethod}%
  \BibitemOpen
  \bibfield  {author} {\bibinfo {author} {\bibfnamefont {N.~F.}\ \bibnamefont
  {Ramsey}},\ }\bibfield  {title} {\bibinfo {title} {A new molecular beam
  resonance method},\ }\href {https://doi.org/10.1103/PhysRev.76.996}
  {\bibfield  {journal} {\bibinfo  {journal} {Phys. Rev.}\ }\textbf {\bibinfo
  {volume} {76}},\ \bibinfo {pages} {996} (\bibinfo {year} {1949})}\BibitemShut
  {NoStop}%
\bibitem [{\citenamefont {Ramsey}(1985)}]{Ramsey1985MolecularBeams}%
  \BibitemOpen
  \bibfield  {author} {\bibinfo {author} {\bibfnamefont {N.}~\bibnamefont
  {Ramsey}},\ }\href {https://books.google.com.sg/books?id=T\_7Hg08X7CMC}
  {\emph {\bibinfo {title} {Molecular Beams}}},\ International series of
  monographs on physics\ (\bibinfo  {publisher} {OUP Oxford},\ \bibinfo {year}
  {1985})\BibitemShut {NoStop}%
\bibitem [{\citenamefont {Cetina}\ \emph {et~al.}(2016)\citenamefont {Cetina},
  \citenamefont {Jag}, \citenamefont {Lous}, \citenamefont {Fritsche},
  \citenamefont {Walraven}, \citenamefont {Grimm}, \citenamefont {Levinsen},
  \citenamefont {Parish}, \citenamefont {Schmidt}, \citenamefont {Knap},\ and\
  \citenamefont {Demler}}]{Cetina2016ManyBodyInterferometry}%
  \BibitemOpen
  \bibfield  {author} {\bibinfo {author} {\bibfnamefont {M.}~\bibnamefont
  {Cetina}}, \bibinfo {author} {\bibfnamefont {M.}~\bibnamefont {Jag}},
  \bibinfo {author} {\bibfnamefont {R.~S.}\ \bibnamefont {Lous}}, \bibinfo
  {author} {\bibfnamefont {I.}~\bibnamefont {Fritsche}}, \bibinfo {author}
  {\bibfnamefont {J.~T.~M.}\ \bibnamefont {Walraven}}, \bibinfo {author}
  {\bibfnamefont {R.}~\bibnamefont {Grimm}}, \bibinfo {author} {\bibfnamefont
  {J.}~\bibnamefont {Levinsen}}, \bibinfo {author} {\bibfnamefont {M.~M.}\
  \bibnamefont {Parish}}, \bibinfo {author} {\bibfnamefont {R.}~\bibnamefont
  {Schmidt}}, \bibinfo {author} {\bibfnamefont {M.}~\bibnamefont {Knap}},\ and\
  \bibinfo {author} {\bibfnamefont {E.}~\bibnamefont {Demler}},\ }\bibfield
  {title} {\bibinfo {title} {Ultrafast many-body interferometry of impurities
  coupled to a fermi sea},\ }\href {https://doi.org/10.1126/science.aaf5134}
  {\bibfield  {journal} {\bibinfo  {journal} {Science}\ }\textbf {\bibinfo
  {volume} {354}},\ \bibinfo {pages} {96} (\bibinfo {year} {2016})},\ \Eprint
  {https://arxiv.org/abs/https://www.science.org/doi/pdf/10.1126/science.aaf5134}
  {https://www.science.org/doi/pdf/10.1126/science.aaf5134} \BibitemShut
  {NoStop}%
\bibitem [{\citenamefont {Li}\ \emph {et~al.}(2016)\citenamefont {Li},
  \citenamefont {Duca}, \citenamefont {Reitter}, \citenamefont {Grusdt},
  \citenamefont {Demler}, \citenamefont {Endres}, \citenamefont
  {Schleier-Smith}, \citenamefont {Bloch},\ and\ \citenamefont
  {Schneider}}]{Li2016WilsonLine}%
  \BibitemOpen
  \bibfield  {author} {\bibinfo {author} {\bibfnamefont {T.}~\bibnamefont
  {Li}}, \bibinfo {author} {\bibfnamefont {L.}~\bibnamefont {Duca}}, \bibinfo
  {author} {\bibfnamefont {M.}~\bibnamefont {Reitter}}, \bibinfo {author}
  {\bibfnamefont {F.}~\bibnamefont {Grusdt}}, \bibinfo {author} {\bibfnamefont
  {E.}~\bibnamefont {Demler}}, \bibinfo {author} {\bibfnamefont
  {M.}~\bibnamefont {Endres}}, \bibinfo {author} {\bibfnamefont
  {M.}~\bibnamefont {Schleier-Smith}}, \bibinfo {author} {\bibfnamefont
  {I.}~\bibnamefont {Bloch}},\ and\ \bibinfo {author} {\bibfnamefont
  {U.}~\bibnamefont {Schneider}},\ }\bibfield  {title} {\bibinfo {title} {Bloch
  state tomography using wilson lines},\ }\href
  {https://www.science.org/doi/10.1126/science.aad5812} {\bibfield  {journal}
  {\bibinfo  {journal} {Science}\ }\textbf {\bibinfo {volume} {352}},\ \bibinfo
  {pages} {1094} (\bibinfo {year} {2016})}\BibitemShut {NoStop}%
\bibitem [{\citenamefont {Lee}\ \emph {et~al.}(2005)\citenamefont {Lee},
  \citenamefont {Brickman}, \citenamefont {Deslauriers}, \citenamefont
  {Haljan}, \citenamefont {Duan},\ and\ \citenamefont
  {Monroe}}]{Lee2005PhaseControlOfTrappedIonQG}%
  \BibitemOpen
  \bibfield  {author} {\bibinfo {author} {\bibfnamefont {P.~J.}\ \bibnamefont
  {Lee}}, \bibinfo {author} {\bibfnamefont {K.-A.}\ \bibnamefont {Brickman}},
  \bibinfo {author} {\bibfnamefont {L.}~\bibnamefont {Deslauriers}}, \bibinfo
  {author} {\bibfnamefont {P.~C.}\ \bibnamefont {Haljan}}, \bibinfo {author}
  {\bibfnamefont {L.-M.}\ \bibnamefont {Duan}},\ and\ \bibinfo {author}
  {\bibfnamefont {C.}~\bibnamefont {Monroe}},\ }\bibfield  {title} {\bibinfo
  {title} {Phase control of trapped ion quantum gates},\ }\href
  {https://doi.org/10.1088/1464-4266/7/10/025} {\bibfield  {journal} {\bibinfo
  {journal} {Journal of Optics B: Quantum and Semiclassical Optics}\ }\textbf
  {\bibinfo {volume} {7}},\ \bibinfo {pages} {S371} (\bibinfo {year}
  {2005})}\BibitemShut {NoStop}%
\bibitem [{\citenamefont {Hu}\ \emph {et~al.}(2018)\citenamefont {Hu},
  \citenamefont {Niu}, \citenamefont {Jin}, \citenamefont {Chen}, \citenamefont
  {Dong}, \citenamefont {Schmiedmayer},\ and\ \citenamefont
  {Zhou}}]{Dong2018RIofMotionalQuantumStates}%
  \BibitemOpen
  \bibfield  {author} {\bibinfo {author} {\bibfnamefont {D.}~\bibnamefont
  {Hu}}, \bibinfo {author} {\bibfnamefont {L.}~\bibnamefont {Niu}}, \bibinfo
  {author} {\bibfnamefont {S.}~\bibnamefont {Jin}}, \bibinfo {author}
  {\bibfnamefont {X.}~\bibnamefont {Chen}}, \bibinfo {author} {\bibfnamefont
  {G.}~\bibnamefont {Dong}}, \bibinfo {author} {\bibfnamefont {J.}~\bibnamefont
  {Schmiedmayer}},\ and\ \bibinfo {author} {\bibfnamefont {X.}~\bibnamefont
  {Zhou}},\ }\bibfield  {title} {\bibinfo {title} {Ramsey interferometry with
  trapped motional quantum states},\ }\href
  {https://doi.org/10.1038/s42005-018-0030-7} {\bibfield  {journal} {\bibinfo
  {journal} {Communications Physics}\ }\textbf {\bibinfo {volume} {1}},\
  \bibinfo {pages} {29} (\bibinfo {year} {2018})}\BibitemShut {NoStop}%
\bibitem [{\citenamefont {Santarelli}\ \emph {et~al.}(1999)\citenamefont
  {Santarelli}, \citenamefont {Laurent}, \citenamefont {Lemonde}, \citenamefont
  {Clairon}, \citenamefont {Mann}, \citenamefont {Chang}, \citenamefont
  {Luiten},\ and\ \citenamefont {Salomon}}]{Santerelli1999QPN}%
  \BibitemOpen
  \bibfield  {author} {\bibinfo {author} {\bibfnamefont {G.}~\bibnamefont
  {Santarelli}}, \bibinfo {author} {\bibfnamefont {P.}~\bibnamefont {Laurent}},
  \bibinfo {author} {\bibfnamefont {P.}~\bibnamefont {Lemonde}}, \bibinfo
  {author} {\bibfnamefont {A.}~\bibnamefont {Clairon}}, \bibinfo {author}
  {\bibfnamefont {A.~G.}\ \bibnamefont {Mann}}, \bibinfo {author}
  {\bibfnamefont {S.}~\bibnamefont {Chang}}, \bibinfo {author} {\bibfnamefont
  {A.~N.}\ \bibnamefont {Luiten}},\ and\ \bibinfo {author} {\bibfnamefont
  {C.}~\bibnamefont {Salomon}},\ }\bibfield  {title} {\bibinfo {title} {Quantum
  projection noise in an atomic fountain: A high stability cesium frequency
  standard},\ }\href {https://doi.org/10.1103/PhysRevLett.82.4619} {\bibfield
  {journal} {\bibinfo  {journal} {Phys. Rev. Lett.}\ }\textbf {\bibinfo
  {volume} {82}},\ \bibinfo {pages} {4619} (\bibinfo {year}
  {1999})}\BibitemShut {NoStop}%
\bibitem [{\citenamefont {Pedrozo-Pe\~nafiel}\ \emph
  {et~al.}(2020)\citenamefont {Pedrozo-Pe\~nafiel}, \citenamefont {Colombo},
  \citenamefont {Shu}, \citenamefont {Adiyatullin}, \citenamefont {Li},
  \citenamefont {Mendez}, \citenamefont {Braverman}, \citenamefont {Kawasaki},
  \citenamefont {Akamatsu}, \citenamefont {Xiao},\ and\ \citenamefont
  {Vuleti\'c}}]{Pedrozo2020EntanglementOnClockTransition}%
  \BibitemOpen
  \bibfield  {author} {\bibinfo {author} {\bibfnamefont {E.}~\bibnamefont
  {Pedrozo-Pe\~nafiel}}, \bibinfo {author} {\bibfnamefont {S.}~\bibnamefont
  {Colombo}}, \bibinfo {author} {\bibfnamefont {C.}~\bibnamefont {Shu}},
  \bibinfo {author} {\bibfnamefont {A.~F.}\ \bibnamefont {Adiyatullin}},
  \bibinfo {author} {\bibfnamefont {Z.}~\bibnamefont {Li}}, \bibinfo {author}
  {\bibfnamefont {E.}~\bibnamefont {Mendez}}, \bibinfo {author} {\bibfnamefont
  {B.}~\bibnamefont {Braverman}}, \bibinfo {author} {\bibfnamefont
  {A.}~\bibnamefont {Kawasaki}}, \bibinfo {author} {\bibfnamefont
  {D.}~\bibnamefont {Akamatsu}}, \bibinfo {author} {\bibfnamefont
  {Y.}~\bibnamefont {Xiao}},\ and\ \bibinfo {author} {\bibfnamefont
  {V.}~\bibnamefont {Vuleti\'c}},\ }\bibfield  {title} {\bibinfo {title}
  {Entanglement on an optical atomic-clock transition},\ }\href
  {https://doi.org/10.1038/s41586-020-3006-1} {\bibfield  {journal} {\bibinfo
  {journal} {Nature}\ }\textbf {\bibinfo {volume} {588}},\ \bibinfo {pages}
  {414} (\bibinfo {year} {2020})}\BibitemShut {NoStop}%
\bibitem [{\citenamefont {Dum}\ and\ \citenamefont
  {Olshanii}(1996)}]{Dum1996GaugeStructuresInALI}%
  \BibitemOpen
  \bibfield  {author} {\bibinfo {author} {\bibfnamefont {R.}~\bibnamefont
  {Dum}}\ and\ \bibinfo {author} {\bibfnamefont {M.}~\bibnamefont {Olshanii}},\
  }\bibfield  {title} {\bibinfo {title} {Gauge structures in atom-laser
  interaction: Bloch oscillations in a dark lattice},\ }\href
  {https://doi.org/10.1103/PhysRevLett.76.1788} {\bibfield  {journal} {\bibinfo
   {journal} {Phys. Rev. Lett.}\ }\textbf {\bibinfo {volume} {76}},\ \bibinfo
  {pages} {1788} (\bibinfo {year} {1996})}\BibitemShut {NoStop}%
\bibitem [{\citenamefont {Dutta}\ \emph {et~al.}(1999)\citenamefont {Dutta},
  \citenamefont {Teo},\ and\ \citenamefont {Raithel}}]{Dutta1999ScalarTerm}%
  \BibitemOpen
  \bibfield  {author} {\bibinfo {author} {\bibfnamefont {S.~K.}\ \bibnamefont
  {Dutta}}, \bibinfo {author} {\bibfnamefont {B.~K.}\ \bibnamefont {Teo}},\
  and\ \bibinfo {author} {\bibfnamefont {G.}~\bibnamefont {Raithel}},\
  }\bibfield  {title} {\bibinfo {title} {Tunneling dynamics and gauge
  potentials in optical lattices},\ }\href
  {https://doi.org/10.1103/PhysRevLett.83.1934} {\bibfield  {journal} {\bibinfo
   {journal} {Phys. Rev. Lett.}\ }\textbf {\bibinfo {volume} {83}},\ \bibinfo
  {pages} {1934} (\bibinfo {year} {1999})}\BibitemShut {NoStop}%
\bibitem [{\citenamefont {Anderson}\ \emph {et~al.}(2020)\citenamefont
  {Anderson}, \citenamefont {Trypogeorgos}, \citenamefont {Vald{\'e}s-Curiel},
  \citenamefont {Liang}, \citenamefont {Tao}, \citenamefont {Zhao},
  \citenamefont {Andrijauskas}, \citenamefont {Juzeli{\=u}nas},\ and\
  \citenamefont {Spielman}}]{Anderson2020SubwavelengthOpticalLattice}%
  \BibitemOpen
  \bibfield  {author} {\bibinfo {author} {\bibfnamefont {R.~P.}\ \bibnamefont
  {Anderson}}, \bibinfo {author} {\bibfnamefont {D.}~\bibnamefont
  {Trypogeorgos}}, \bibinfo {author} {\bibfnamefont {A.}~\bibnamefont
  {Vald{\'e}s-Curiel}}, \bibinfo {author} {\bibfnamefont {Q.-Y.}\ \bibnamefont
  {Liang}}, \bibinfo {author} {\bibfnamefont {J.}~\bibnamefont {Tao}}, \bibinfo
  {author} {\bibfnamefont {M.}~\bibnamefont {Zhao}}, \bibinfo {author}
  {\bibfnamefont {T.}~\bibnamefont {Andrijauskas}}, \bibinfo {author}
  {\bibfnamefont {G.}~\bibnamefont {Juzeli{\=u}nas}},\ and\ \bibinfo {author}
  {\bibfnamefont {I.}~\bibnamefont {Spielman}},\ }\bibfield  {title} {\bibinfo
  {title} {Realization of a deeply subwavelength adiabatic optical lattice},\
  }\href {https://doi.org/10.1103/PhysRevResearch.2.013149} {\bibfield
  {journal} {\bibinfo  {journal} {Phys. Rev. Res.}\ }\textbf {\bibinfo {volume}
  {2}},\ \bibinfo {pages} {013149} (\bibinfo {year} {2020})}\BibitemShut
  {NoStop}%
\bibitem [{\citenamefont {Gvozdiovas}\ \emph {et~al.}(2021)\citenamefont
  {Gvozdiovas}, \citenamefont {Rackauskas},\ and\ \citenamefont
  {Juzeli{\=u}nas}}]{10.21468/SciPostPhys.11.6.100}%
  \BibitemOpen
  \bibfield  {author} {\bibinfo {author} {\bibfnamefont {E.}~\bibnamefont
  {Gvozdiovas}}, \bibinfo {author} {\bibfnamefont {P.}~\bibnamefont
  {Rackauskas}},\ and\ \bibinfo {author} {\bibfnamefont {G.}~\bibnamefont
  {Juzeli{\=u}nas}},\ }\bibfield  {title} {\bibinfo {title} {Optical lattice
  with spin-dependent sub-wavelength barriers},\ }\href
  {https://doi.org/10.21468/SciPostPhys.11.6.100} {\bibfield  {journal}
  {\bibinfo  {journal} {SciPost Phys.}\ }\textbf {\bibinfo {volume} {11}},\
  \bibinfo {pages} {100} (\bibinfo {year} {2021})}\BibitemShut {NoStop}%
\bibitem [{\citenamefont {Wang}\ \emph {et~al.}(2018)\citenamefont {Wang},
  \citenamefont {Subhankar}, \citenamefont {Bienias}, \citenamefont {Lacki},
  \citenamefont {Tsui}, \citenamefont {Baranov}, \citenamefont {Gorshkov},
  \citenamefont {Zoller}, \citenamefont {Porto},\ and\ \citenamefont
  {Rolston}}]{Wang2018DarkStateOpticalLattice}%
  \BibitemOpen
  \bibfield  {author} {\bibinfo {author} {\bibfnamefont {Y.}~\bibnamefont
  {Wang}}, \bibinfo {author} {\bibfnamefont {S.}~\bibnamefont {Subhankar}},
  \bibinfo {author} {\bibfnamefont {P.}~\bibnamefont {Bienias}}, \bibinfo
  {author} {\bibfnamefont {M.}~\bibnamefont {Lacki}}, \bibinfo {author}
  {\bibfnamefont {T.-C.}\ \bibnamefont {Tsui}}, \bibinfo {author}
  {\bibfnamefont {M.~A.}\ \bibnamefont {Baranov}}, \bibinfo {author}
  {\bibfnamefont {A.~V.}\ \bibnamefont {Gorshkov}}, \bibinfo {author}
  {\bibfnamefont {P.}~\bibnamefont {Zoller}}, \bibinfo {author} {\bibfnamefont
  {J.~V.}\ \bibnamefont {Porto}},\ and\ \bibinfo {author} {\bibfnamefont
  {S.~L.}\ \bibnamefont {Rolston}},\ }\bibfield  {title} {\bibinfo {title}
  {Dark state optical lattice with a subwavelength spatial structure},\ }\href
  {https://doi.org/10.1103/PhysRevLett.120.083601} {\bibfield  {journal}
  {\bibinfo  {journal} {Phys. Rev. Lett.}\ }\textbf {\bibinfo {volume} {120}},\
  \bibinfo {pages} {083601} (\bibinfo {year} {2018})}\BibitemShut {NoStop}%
\bibitem [{\citenamefont {Dalibard}\ \emph {et~al.}(2011)\citenamefont
  {Dalibard}, \citenamefont {Gerbier}, \citenamefont {Juzeli{\=u}nas},\ and\
  \citenamefont {{\"O}hberg}}]{dalibard2011colloquium}%
  \BibitemOpen
  \bibfield  {author} {\bibinfo {author} {\bibfnamefont {J.}~\bibnamefont
  {Dalibard}}, \bibinfo {author} {\bibfnamefont {F.}~\bibnamefont {Gerbier}},
  \bibinfo {author} {\bibfnamefont {G.}~\bibnamefont {Juzeli{\=u}nas}},\ and\
  \bibinfo {author} {\bibfnamefont {P.}~\bibnamefont {{\"O}hberg}},\ }\bibfield
   {title} {\bibinfo {title} {Colloquium: Artificial gauge potentials for
  neutral atoms},\ }\href {https://doi.org/10.1103/RevModPhys.83.1523}
  {\bibfield  {journal} {\bibinfo  {journal} {Reviews of Modern Physics}\
  }\textbf {\bibinfo {volume} {83}},\ \bibinfo {pages} {1523} (\bibinfo {year}
  {2011})}\BibitemShut {NoStop}%
\bibitem [{\citenamefont {Toyoda}\ \emph {et~al.}(2013)\citenamefont {Toyoda},
  \citenamefont {Uchida}, \citenamefont {Noguchi}, \citenamefont {Haze},\ and\
  \citenamefont {Urabe}}]{Toyoda2013HolonomicSingleQubitOperations}%
  \BibitemOpen
  \bibfield  {author} {\bibinfo {author} {\bibfnamefont {K.}~\bibnamefont
  {Toyoda}}, \bibinfo {author} {\bibfnamefont {K.}~\bibnamefont {Uchida}},
  \bibinfo {author} {\bibfnamefont {A.}~\bibnamefont {Noguchi}}, \bibinfo
  {author} {\bibfnamefont {S.}~\bibnamefont {Haze}},\ and\ \bibinfo {author}
  {\bibfnamefont {S.}~\bibnamefont {Urabe}},\ }\bibfield  {title} {\bibinfo
  {title} {Realization of holonomic single-qubit operations},\ }\href
  {https://doi.org/10.1103/PhysRevA.87.052307} {\bibfield  {journal} {\bibinfo
  {journal} {Phys. Rev. A}\ }\textbf {\bibinfo {volume} {87}},\ \bibinfo
  {pages} {052307} (\bibinfo {year} {2013})}\BibitemShut {NoStop}%
\bibitem [{\citenamefont {Chaneliere}\ \emph {et~al.}(2008)\citenamefont
  {Chaneliere}, \citenamefont {He}, \citenamefont {Kaiser},\ and\ \citenamefont
  {Wilkowski}}]{chaneliere2008three}%
  \BibitemOpen
  \bibfield  {author} {\bibinfo {author} {\bibfnamefont {T.}~\bibnamefont
  {Chaneliere}}, \bibinfo {author} {\bibfnamefont {L.}~\bibnamefont {He}},
  \bibinfo {author} {\bibfnamefont {R.}~\bibnamefont {Kaiser}},\ and\ \bibinfo
  {author} {\bibfnamefont {D.}~\bibnamefont {Wilkowski}},\ }\bibfield  {title}
  {\bibinfo {title} {Three dimensional cooling and trapping with a narrow
  line},\ }\href {https://link.springer.com/article/10.1140/epjd/e2007-00329-8}
  {\bibfield  {journal} {\bibinfo  {journal} {The European Physical Journal D}\
  }\textbf {\bibinfo {volume} {46}},\ \bibinfo {pages} {507} (\bibinfo {year}
  {2008})}\BibitemShut {NoStop}%
\bibitem [{\citenamefont {Yang}\ \emph {et~al.}(2015)\citenamefont {Yang},
  \citenamefont {Pandey}, \citenamefont {Pramod}, \citenamefont {Leroux},
  \citenamefont {Kwong}, \citenamefont {Hajiyev}, \citenamefont {Chia},
  \citenamefont {Fang},\ and\ \citenamefont {Wilkowski}}]{Tao2015HighFlux}%
  \BibitemOpen
  \bibfield  {author} {\bibinfo {author} {\bibfnamefont {T.}~\bibnamefont
  {Yang}}, \bibinfo {author} {\bibfnamefont {K.}~\bibnamefont {Pandey}},
  \bibinfo {author} {\bibfnamefont {M.~S.}\ \bibnamefont {Pramod}}, \bibinfo
  {author} {\bibfnamefont {F.}~\bibnamefont {Leroux}}, \bibinfo {author}
  {\bibfnamefont {C.~C.}\ \bibnamefont {Kwong}}, \bibinfo {author}
  {\bibfnamefont {E.}~\bibnamefont {Hajiyev}}, \bibinfo {author} {\bibfnamefont
  {Z.~Y.}\ \bibnamefont {Chia}}, \bibinfo {author} {\bibfnamefont
  {B.}~\bibnamefont {Fang}},\ and\ \bibinfo {author} {\bibfnamefont
  {D.}~\bibnamefont {Wilkowski}},\ }\bibfield  {title} {\bibinfo {title} {A
  high flux source of cold strontium atoms},\ }\href
  {https://doi.org/10.1140/epjd/e2015-60288-y} {\bibfield  {journal} {\bibinfo
  {journal} {The European Physical Journal D}\ }\textbf {\bibinfo {volume}
  {69}},\ \bibinfo {pages} {226} (\bibinfo {year} {2015})}\BibitemShut
  {NoStop}%
\bibitem [{\citenamefont {Hasan}\ \emph
  {et~al.}(2022{\natexlab{a}})\citenamefont {Hasan}, \citenamefont {Madasu},
  \citenamefont {Rathod}, \citenamefont {Kwong}, \citenamefont {Miniatura},
  \citenamefont {Chevy},\ and\ \citenamefont
  {Wilkowski}}]{hasan2022anisotropic}%
  \BibitemOpen
  \bibfield  {author} {\bibinfo {author} {\bibfnamefont {M.}~\bibnamefont
  {Hasan}}, \bibinfo {author} {\bibfnamefont {C.~S.}\ \bibnamefont {Madasu}},
  \bibinfo {author} {\bibfnamefont {K.~D.}\ \bibnamefont {Rathod}}, \bibinfo
  {author} {\bibfnamefont {C.~C.}\ \bibnamefont {Kwong}}, \bibinfo {author}
  {\bibfnamefont {C.}~\bibnamefont {Miniatura}}, \bibinfo {author}
  {\bibfnamefont {F.}~\bibnamefont {Chevy}},\ and\ \bibinfo {author}
  {\bibfnamefont {D.}~\bibnamefont {Wilkowski}},\ }\bibfield  {title} {\bibinfo
  {title} {Wave packet dynamics in synthetic non-abelian gauge fields},\ }\href
  {https://doi.org/10.1103/PhysRevLett.129.130402} {\bibfield  {journal}
  {\bibinfo  {journal} {Phys. Rev. Lett.}\ }\textbf {\bibinfo {volume} {129}},\
  \bibinfo {pages} {130402} (\bibinfo {year} {2022}{\natexlab{a}})}\BibitemShut
  {NoStop}%
\bibitem [{\citenamefont {Hasan}\ \emph
  {et~al.}(2022{\natexlab{b}})\citenamefont {Hasan}, \citenamefont {Madasu},
  \citenamefont {Rathod}, \citenamefont {Kwong},\ and\ \citenamefont
  {Wilkowski}}]{Hasan_2022}%
  \BibitemOpen
  \bibfield  {author} {\bibinfo {author} {\bibfnamefont {M.}~\bibnamefont
  {Hasan}}, \bibinfo {author} {\bibfnamefont {C.}~\bibnamefont {Madasu}},
  \bibinfo {author} {\bibfnamefont {K.}~\bibnamefont {Rathod}}, \bibinfo
  {author} {\bibfnamefont {C.}~\bibnamefont {Kwong}},\ and\ \bibinfo {author}
  {\bibfnamefont {D.}~\bibnamefont {Wilkowski}},\ }\bibfield  {title} {\bibinfo
  {title} {Evolution of an ultracold gas in a non-abelian gauge field: finite
  temperature effect},\ }\href {https://doi.org/10.1070/qel18071} {\bibfield
  {journal} {\bibinfo  {journal} {Quantum Electronics}\ }\textbf {\bibinfo
  {volume} {52}},\ \bibinfo {pages} {532} (\bibinfo {year}
  {2022}{\natexlab{b}})}\BibitemShut {NoStop}%
\bibitem [{\citenamefont {Chetan Sriram~et. al.}()}]{SM}%
  \BibitemOpen
  \bibfield  {author} {\bibinfo {author} {\bibfnamefont {M.}~\bibnamefont
  {Chetan Sriram~et. al.}},\ }\bibfield  {title} {\bibinfo {title} {See
  supplemental material at for a detailed description of sample preparation,
  detection procedure, adiabatic approximation, and other experimental
  considerations},\ }\href
  {http://link.aps.org/supplemental/10.1103/PhysRevApplied.21.L051001} {\
  }\BibitemShut {NoStop}%
\bibitem [{\citenamefont {Madasu}\ \emph {et~al.}(2022)\citenamefont {Madasu},
  \citenamefont {Hasan}, \citenamefont {Rathod}, \citenamefont {Kwong},\ and\
  \citenamefont {Wilkowski}}]{Chetan2022AtomtronicDDT}%
  \BibitemOpen
  \bibfield  {author} {\bibinfo {author} {\bibfnamefont {C.~S.}\ \bibnamefont
  {Madasu}}, \bibinfo {author} {\bibfnamefont {M.}~\bibnamefont {Hasan}},
  \bibinfo {author} {\bibfnamefont {K.~D.}\ \bibnamefont {Rathod}}, \bibinfo
  {author} {\bibfnamefont {C.~C.}\ \bibnamefont {Kwong}},\ and\ \bibinfo
  {author} {\bibfnamefont {D.}~\bibnamefont {Wilkowski}},\ }\bibfield  {title}
  {\bibinfo {title} {Datta-das transistor for atomtronic circuits using
  artificial gauge fields},\ }\href
  {https://doi.org/10.1103/PhysRevResearch.4.033180} {\bibfield  {journal}
  {\bibinfo  {journal} {Phys. Rev. Research}\ }\textbf {\bibinfo {volume}
  {4}},\ \bibinfo {pages} {033180} (\bibinfo {year} {2022})}\BibitemShut
  {NoStop}%
\bibitem [{\citenamefont {Vitanov}\ \emph {et~al.}(2017)\citenamefont
  {Vitanov}, \citenamefont {Rangelov}, \citenamefont {Shore},\ and\
  \citenamefont {Bergmann}}]{Vitanov2017STIRAPReview}%
  \BibitemOpen
  \bibfield  {author} {\bibinfo {author} {\bibfnamefont {N.~V.}\ \bibnamefont
  {Vitanov}}, \bibinfo {author} {\bibfnamefont {A.~A.}\ \bibnamefont
  {Rangelov}}, \bibinfo {author} {\bibfnamefont {B.~W.}\ \bibnamefont
  {Shore}},\ and\ \bibinfo {author} {\bibfnamefont {K.}~\bibnamefont
  {Bergmann}},\ }\bibfield  {title} {\bibinfo {title} {Stimulated raman
  adiabatic passage in physics, chemistry, and beyond},\ }\href
  {https://doi.org/10.1103/RevModPhys.89.015006} {\bibfield  {journal}
  {\bibinfo  {journal} {Reviews of Modern Physics}\ }\textbf {\bibinfo {volume}
  {89}},\ \bibinfo {pages} {015006} (\bibinfo {year} {2017})}\BibitemShut
  {NoStop}%
\bibitem [{\citenamefont {Ammann}\ and\ \citenamefont
  {Christensen}(1997)}]{PhysRevLett.78.2088}%
  \BibitemOpen
  \bibfield  {author} {\bibinfo {author} {\bibfnamefont {H.}~\bibnamefont
  {Ammann}}\ and\ \bibinfo {author} {\bibfnamefont {N.}~\bibnamefont
  {Christensen}},\ }\bibfield  {title} {\bibinfo {title} {Delta kick cooling: A
  new method for cooling atoms},\ }\href
  {https://doi.org/10.1103/PhysRevLett.78.2088} {\bibfield  {journal} {\bibinfo
   {journal} {Phys. Rev. Lett.}\ }\textbf {\bibinfo {volume} {78}},\ \bibinfo
  {pages} {2088} (\bibinfo {year} {1997})}\BibitemShut {NoStop}%
\bibitem [{\citenamefont {Vald{\'e}s-Curiel}\ \emph {et~al.}(2021)\citenamefont
  {Vald{\'e}s-Curiel}, \citenamefont {Trypogeorgos}, \citenamefont {Liang},
  \citenamefont {Anderson},\ and\ \citenamefont
  {Spielman}}]{valdes-curiel_topological_2021}%
  \BibitemOpen
  \bibfield  {author} {\bibinfo {author} {\bibfnamefont {A.}~\bibnamefont
  {Vald{\'e}s-Curiel}}, \bibinfo {author} {\bibfnamefont {D.}~\bibnamefont
  {Trypogeorgos}}, \bibinfo {author} {\bibfnamefont {Q.-Y.}\ \bibnamefont
  {Liang}}, \bibinfo {author} {\bibfnamefont {R.~P.}\ \bibnamefont
  {Anderson}},\ and\ \bibinfo {author} {\bibfnamefont {I.}~\bibnamefont
  {Spielman}},\ }\bibfield  {title} {\bibinfo {title} {Topological features
  without a lattice in rashba spin-orbit coupled atoms},\ }\href
  {https://doi.org/10.1038/s41467-020-20762-4} {\bibfield  {journal} {\bibinfo
  {journal} {Nature Communications}\ }\textbf {\bibinfo {volume} {12}},\
  \bibinfo {pages} {593} (\bibinfo {year} {2021})}\BibitemShut {NoStop}%
\bibitem [{\citenamefont {Leroux}\ \emph {et~al.}(2018)\citenamefont {Leroux},
  \citenamefont {Pandey}, \citenamefont {Rehbi}, \citenamefont {Chevy},
  \citenamefont {Miniatura}, \citenamefont {Gr{\'e}maud},\ and\ \citenamefont
  {Wilkowski}}]{Frederic2018Non-Abelian}%
  \BibitemOpen
  \bibfield  {author} {\bibinfo {author} {\bibfnamefont {F.}~\bibnamefont
  {Leroux}}, \bibinfo {author} {\bibfnamefont {K.}~\bibnamefont {Pandey}},
  \bibinfo {author} {\bibfnamefont {R.}~\bibnamefont {Rehbi}}, \bibinfo
  {author} {\bibfnamefont {F.}~\bibnamefont {Chevy}}, \bibinfo {author}
  {\bibfnamefont {C.}~\bibnamefont {Miniatura}}, \bibinfo {author}
  {\bibfnamefont {B.}~\bibnamefont {Gr{\'e}maud}},\ and\ \bibinfo {author}
  {\bibfnamefont {D.}~\bibnamefont {Wilkowski}},\ }\bibfield  {title} {\bibinfo
  {title} {Non-abelian adiabatic geometric transformations in a cold strontium
  gas},\ }\href {https://doi.org/10.1038/s41467-018-05865-3} {\bibfield
  {journal} {\bibinfo  {journal} {Nature Communications}\ }\textbf {\bibinfo
  {volume} {9}},\ \bibinfo {pages} {3580} (\bibinfo {year} {2018})}\BibitemShut
  {NoStop}%
\bibitem [{\citenamefont {Ruseckas}\ \emph {et~al.}(2005)\citenamefont
  {Ruseckas}, \citenamefont {Juzeli{\=u}nas}, \citenamefont {{\"O}hberg},\ and\
  \citenamefont {Fleischhauer}}]{Ruseckas2005Non-Abelian}%
  \BibitemOpen
  \bibfield  {author} {\bibinfo {author} {\bibfnamefont {J.}~\bibnamefont
  {Ruseckas}}, \bibinfo {author} {\bibfnamefont {G.}~\bibnamefont
  {Juzeli{\=u}nas}}, \bibinfo {author} {\bibfnamefont {P.}~\bibnamefont
  {{\"O}hberg}},\ and\ \bibinfo {author} {\bibfnamefont {M.}~\bibnamefont
  {Fleischhauer}},\ }\bibfield  {title} {\bibinfo {title} {Non-abelian gauge
  potentials for ultracold atoms with degenerate dark states},\ }\href
  {https://doi.org/10.1103/PhysRevLett.95.010404} {\bibfield  {journal}
  {\bibinfo  {journal} {Phys. Rev. Lett.}\ }\textbf {\bibinfo {volume} {95}},\
  \bibinfo {pages} {010404} (\bibinfo {year} {2005})}\BibitemShut {NoStop}%
\bibitem [{\citenamefont {Vaishnav}\ \emph {et~al.}(2008)\citenamefont
  {Vaishnav}, \citenamefont {Ruseckas}, \citenamefont {Clark},\ and\
  \citenamefont {Juzeli{\=u}nas}}]{Vaishnav2008DDT}%
  \BibitemOpen
  \bibfield  {author} {\bibinfo {author} {\bibfnamefont {J.~Y.}\ \bibnamefont
  {Vaishnav}}, \bibinfo {author} {\bibfnamefont {J.}~\bibnamefont {Ruseckas}},
  \bibinfo {author} {\bibfnamefont {C.~W.}\ \bibnamefont {Clark}},\ and\
  \bibinfo {author} {\bibfnamefont {G.}~\bibnamefont {Juzeli{\=u}nas}},\
  }\bibfield  {title} {\bibinfo {title} {Spin field effect transistors with
  ultracold atoms},\ }\href {https://doi.org/10.1103/PhysRevLett.101.265302}
  {\bibfield  {journal} {\bibinfo  {journal} {Physical Review Letters}\
  }\textbf {\bibinfo {volume} {101}},\ \bibinfo {pages} {265302} (\bibinfo
  {year} {2008})}\BibitemShut {NoStop}%
\bibitem [{\citenamefont {Kubala}\ \emph {et~al.}(2021)\citenamefont {Kubala},
  \citenamefont {Zakrzewski},\ and\ \citenamefont
  {Lacki}}]{Kubala2021OpticalLatticeForTripod}%
  \BibitemOpen
  \bibfield  {author} {\bibinfo {author} {\bibfnamefont {P.}~\bibnamefont
  {Kubala}}, \bibinfo {author} {\bibfnamefont {J.}~\bibnamefont {Zakrzewski}},\
  and\ \bibinfo {author} {\bibfnamefont {M.}~\bibnamefont {Lacki}},\ }\bibfield
   {title} {\bibinfo {title} {Optical lattice for a tripodlike atomic level
  structure},\ }\href {https://doi.org/10.1103/PhysRevA.104.053312} {\bibfield
  {journal} {\bibinfo  {journal} {Phys. Rev. A}\ }\textbf {\bibinfo {volume}
  {104}},\ \bibinfo {pages} {053312} (\bibinfo {year} {2021})}\BibitemShut
  {NoStop}%
\bibitem [{\citenamefont {Tsui}\ \emph {et~al.}(2020)\citenamefont {Tsui},
  \citenamefont {Wang}, \citenamefont {Subhankar}, \citenamefont {Porto},\ and\
  \citenamefont {Rolston}}]{Tsui2020StroboscopicOL}%
  \BibitemOpen
  \bibfield  {author} {\bibinfo {author} {\bibfnamefont {T.-C.}\ \bibnamefont
  {Tsui}}, \bibinfo {author} {\bibfnamefont {Y.}~\bibnamefont {Wang}}, \bibinfo
  {author} {\bibfnamefont {S.}~\bibnamefont {Subhankar}}, \bibinfo {author}
  {\bibfnamefont {J.~V.}\ \bibnamefont {Porto}},\ and\ \bibinfo {author}
  {\bibfnamefont {S.~L.}\ \bibnamefont {Rolston}},\ }\bibfield  {title}
  {\bibinfo {title} {Realization of a stroboscopic optical lattice for cold
  atoms with subwavelength spacing},\ }\href
  {https://doi.org/10.1103/PhysRevA.101.041603} {\bibfield  {journal} {\bibinfo
   {journal} {Phys. Rev. A}\ }\textbf {\bibinfo {volume} {101}},\ \bibinfo
  {pages} {041603} (\bibinfo {year} {2020})}\BibitemShut {NoStop}%
\bibitem [{\citenamefont {Frazer}\ and\ \citenamefont
  {Gillen}(2022)}]{Frazer20221DArrays}%
  \BibitemOpen
  \bibfield  {author} {\bibinfo {author} {\bibfnamefont {T.}~\bibnamefont
  {Frazer}}\ and\ \bibinfo {author} {\bibfnamefont {K.}~\bibnamefont
  {Gillen}},\ }\bibfield  {title} {\bibinfo {title} {One-dimensional arrays of
  optical dark spot traps from nested gaussian laser beams for quantum
  computing},\ }\href
  {https://link.springer.com/article/10.1007/s00340-022-07808-9} {\bibfield
  {journal} {\bibinfo  {journal} {Applied Physics B}\ }\textbf {\bibinfo
  {volume} {128}},\ \bibinfo {pages} {90} (\bibinfo {year} {2022})}\BibitemShut
  {NoStop}%
\bibitem [{\citenamefont {Amico}\ \emph {et~al.}(2021)\citenamefont {Amico},
  \citenamefont {Boshier}, \citenamefont {Birkl}, \citenamefont {Minguzzi},
  \citenamefont {Miniatura}, \citenamefont {Kwek}, \citenamefont {Aghamalyan},
  \citenamefont {Ahufinger}, \citenamefont {Anderson}, \citenamefont {Andrei},
  \citenamefont {Arnold}, \citenamefont {Baker}, \citenamefont {Bell},
  \citenamefont {Bland}, \citenamefont {Brantut}, \citenamefont {Cassettari},
  \citenamefont {Chetcuti}, \citenamefont {Chevy}, \citenamefont {Citro},
  \citenamefont {De~Palo}, \citenamefont {Dumke}, \citenamefont {Edwards},
  \citenamefont {Folman}, \citenamefont {Fortagh}, \citenamefont {Gardiner},
  \citenamefont {Garraway}, \citenamefont {Gauthier}, \citenamefont
  {G{\"u}nther}, \citenamefont {Haug}, \citenamefont {Hufnagel}, \citenamefont
  {Keil}, \citenamefont {Ireland}, \citenamefont {Lebrat}, \citenamefont {Li},
  \citenamefont {Longchambon}, \citenamefont {Mompart}, \citenamefont {Morsch},
  \citenamefont {Naldesi}, \citenamefont {Neely}, \citenamefont {Olshanii},
  \citenamefont {Orignac}, \citenamefont {Pandey}, \citenamefont
  {P{\'e}rez-Obiol}, \citenamefont {Perrin}, \citenamefont {Piroli},
  \citenamefont {Polo}, \citenamefont {Pritchard}, \citenamefont {Proukakis},
  \citenamefont {Rylands}, \citenamefont {Rubinsztein-Dunlop}, \citenamefont
  {Scazza}, \citenamefont {Stringari}, \citenamefont {Tosto}, \citenamefont
  {Trombettoni}, \citenamefont {Victorin}, \citenamefont {Klitzing},
  \citenamefont {Wilkowski}, \citenamefont {Xhani},\ and\ \citenamefont
  {Yakimenko}}]{amico2021roadmap}%
  \BibitemOpen
  \bibfield  {author} {\bibinfo {author} {\bibfnamefont {L.}~\bibnamefont
  {Amico}}, \bibinfo {author} {\bibfnamefont {M.}~\bibnamefont {Boshier}},
  \bibinfo {author} {\bibfnamefont {G.}~\bibnamefont {Birkl}}, \bibinfo
  {author} {\bibfnamefont {A.}~\bibnamefont {Minguzzi}}, \bibinfo {author}
  {\bibfnamefont {C.}~\bibnamefont {Miniatura}}, \bibinfo {author}
  {\bibfnamefont {L.-C.}\ \bibnamefont {Kwek}}, \bibinfo {author}
  {\bibfnamefont {D.}~\bibnamefont {Aghamalyan}}, \bibinfo {author}
  {\bibfnamefont {V.}~\bibnamefont {Ahufinger}}, \bibinfo {author}
  {\bibfnamefont {D.}~\bibnamefont {Anderson}}, \bibinfo {author}
  {\bibfnamefont {N.}~\bibnamefont {Andrei}}, \bibinfo {author} {\bibfnamefont
  {A.~S.}\ \bibnamefont {Arnold}}, \bibinfo {author} {\bibfnamefont
  {M.}~\bibnamefont {Baker}}, \bibinfo {author} {\bibfnamefont {T.~A.}\
  \bibnamefont {Bell}}, \bibinfo {author} {\bibfnamefont {T.}~\bibnamefont
  {Bland}}, \bibinfo {author} {\bibfnamefont {J.~P.}\ \bibnamefont {Brantut}},
  \bibinfo {author} {\bibfnamefont {D.}~\bibnamefont {Cassettari}}, \bibinfo
  {author} {\bibfnamefont {W.~J.}\ \bibnamefont {Chetcuti}}, \bibinfo {author}
  {\bibfnamefont {F.}~\bibnamefont {Chevy}}, \bibinfo {author} {\bibfnamefont
  {R.}~\bibnamefont {Citro}}, \bibinfo {author} {\bibfnamefont
  {S.}~\bibnamefont {De~Palo}}, \bibinfo {author} {\bibfnamefont
  {R.}~\bibnamefont {Dumke}}, \bibinfo {author} {\bibfnamefont
  {M.}~\bibnamefont {Edwards}}, \bibinfo {author} {\bibfnamefont
  {R.}~\bibnamefont {Folman}}, \bibinfo {author} {\bibfnamefont
  {J.}~\bibnamefont {Fortagh}}, \bibinfo {author} {\bibfnamefont {S.~A.}\
  \bibnamefont {Gardiner}}, \bibinfo {author} {\bibfnamefont {B.~M.}\
  \bibnamefont {Garraway}}, \bibinfo {author} {\bibfnamefont {G.}~\bibnamefont
  {Gauthier}}, \bibinfo {author} {\bibfnamefont {A.}~\bibnamefont
  {G{\"u}nther}}, \bibinfo {author} {\bibfnamefont {T.}~\bibnamefont {Haug}},
  \bibinfo {author} {\bibfnamefont {C.}~\bibnamefont {Hufnagel}}, \bibinfo
  {author} {\bibfnamefont {M.}~\bibnamefont {Keil}}, \bibinfo {author}
  {\bibfnamefont {P.}~\bibnamefont {Ireland}}, \bibinfo {author} {\bibfnamefont
  {M.}~\bibnamefont {Lebrat}}, \bibinfo {author} {\bibfnamefont
  {W.}~\bibnamefont {Li}}, \bibinfo {author} {\bibfnamefont {L.}~\bibnamefont
  {Longchambon}}, \bibinfo {author} {\bibfnamefont {J.}~\bibnamefont
  {Mompart}}, \bibinfo {author} {\bibfnamefont {O.}~\bibnamefont {Morsch}},
  \bibinfo {author} {\bibfnamefont {P.}~\bibnamefont {Naldesi}}, \bibinfo
  {author} {\bibfnamefont {T.~W.}\ \bibnamefont {Neely}}, \bibinfo {author}
  {\bibfnamefont {M.}~\bibnamefont {Olshanii}}, \bibinfo {author}
  {\bibfnamefont {E.}~\bibnamefont {Orignac}}, \bibinfo {author} {\bibfnamefont
  {S.}~\bibnamefont {Pandey}}, \bibinfo {author} {\bibfnamefont
  {A.}~\bibnamefont {P{\'e}rez-Obiol}}, \bibinfo {author} {\bibfnamefont
  {H.}~\bibnamefont {Perrin}}, \bibinfo {author} {\bibfnamefont
  {L.}~\bibnamefont {Piroli}}, \bibinfo {author} {\bibfnamefont
  {J.}~\bibnamefont {Polo}}, \bibinfo {author} {\bibfnamefont {A.~L.}\
  \bibnamefont {Pritchard}}, \bibinfo {author} {\bibfnamefont {N.~P.}\
  \bibnamefont {Proukakis}}, \bibinfo {author} {\bibfnamefont {C.}~\bibnamefont
  {Rylands}}, \bibinfo {author} {\bibfnamefont {H.}~\bibnamefont
  {Rubinsztein-Dunlop}}, \bibinfo {author} {\bibfnamefont {F.}~\bibnamefont
  {Scazza}}, \bibinfo {author} {\bibfnamefont {S.}~\bibnamefont {Stringari}},
  \bibinfo {author} {\bibfnamefont {F.}~\bibnamefont {Tosto}}, \bibinfo
  {author} {\bibfnamefont {A.}~\bibnamefont {Trombettoni}}, \bibinfo {author}
  {\bibfnamefont {N.}~\bibnamefont {Victorin}}, \bibinfo {author}
  {\bibfnamefont {W.~v.}\ \bibnamefont {Klitzing}}, \bibinfo {author}
  {\bibfnamefont {D.}~\bibnamefont {Wilkowski}}, \bibinfo {author}
  {\bibfnamefont {K.}~\bibnamefont {Xhani}},\ and\ \bibinfo {author}
  {\bibfnamefont {A.}~\bibnamefont {Yakimenko}},\ }\bibfield  {title} {\bibinfo
  {title} {Roadmap on atomtronics: State of the art and perspective},\ }\href
  {https://doi.org/10.1116/5.0026178} {\bibfield  {journal} {\bibinfo
  {journal} {AVS Quantum Science}\ }\textbf {\bibinfo {volume} {3}},\ \bibinfo
  {pages} {039201} (\bibinfo {year} {2021})},\ \Eprint
  {https://arxiv.org/abs/https://doi.org/10.1116/5.0026178}
  {https://doi.org/10.1116/5.0026178} \BibitemShut {NoStop}%
\bibitem [{\citenamefont {Bord{\'e}}\ \emph {et~al.}(1984)\citenamefont
  {Bord{\'e}}, \citenamefont {Salomon}, \citenamefont {Avrillier},
  \citenamefont {Van~Lerberghe}, \citenamefont {Br{\'e}ant}, \citenamefont
  {Bassi},\ and\ \citenamefont {Scoles}}]{borde1984optical}%
  \BibitemOpen
  \bibfield  {author} {\bibinfo {author} {\bibfnamefont {C.~J.}\ \bibnamefont
  {Bord{\'e}}}, \bibinfo {author} {\bibfnamefont {C.}~\bibnamefont {Salomon}},
  \bibinfo {author} {\bibfnamefont {S.}~\bibnamefont {Avrillier}}, \bibinfo
  {author} {\bibfnamefont {A.}~\bibnamefont {Van~Lerberghe}}, \bibinfo {author}
  {\bibfnamefont {C.}~\bibnamefont {Br{\'e}ant}}, \bibinfo {author}
  {\bibfnamefont {D.}~\bibnamefont {Bassi}},\ and\ \bibinfo {author}
  {\bibfnamefont {G.}~\bibnamefont {Scoles}},\ }\bibfield  {title} {\bibinfo
  {title} {Optical ramsey fringes with traveling waves},\ }\href
  {https://doi.org/10.1103/PhysRevA.30.1836} {\bibfield  {journal} {\bibinfo
  {journal} {Physical Review A}\ }\textbf {\bibinfo {volume} {30}},\ \bibinfo
  {pages} {1836} (\bibinfo {year} {1984})}\BibitemShut {NoStop}%
\bibitem [{\citenamefont {Bord{\'e}}(1989)}]{borde1989atomic}%
  \BibitemOpen
  \bibfield  {author} {\bibinfo {author} {\bibfnamefont {C.~J.}\ \bibnamefont
  {Bord{\'e}}},\ }\bibfield  {title} {\bibinfo {title} {Atomic interferometry
  with internal state labelling},\ }\href
  {https://doi.org/10.1016/0375-9601(89)90537-9} {\bibfield  {journal}
  {\bibinfo  {journal} {Physics letters A}\ }\textbf {\bibinfo {volume}
  {140}},\ \bibinfo {pages} {10} (\bibinfo {year} {1989})}\BibitemShut
  {NoStop}%
\bibitem [{\citenamefont {Kasevich}\ and\ \citenamefont
  {Chu}(1991)}]{PhysRevLett.67.181}%
  \BibitemOpen
  \bibfield  {author} {\bibinfo {author} {\bibfnamefont {M.}~\bibnamefont
  {Kasevich}}\ and\ \bibinfo {author} {\bibfnamefont {S.}~\bibnamefont {Chu}},\
  }\bibfield  {title} {\bibinfo {title} {Atomic interferometry using stimulated
  raman transitions},\ }\href {https://doi.org/10.1103/PhysRevLett.67.181}
  {\bibfield  {journal} {\bibinfo  {journal} {Phys. Rev. Lett.}\ }\textbf
  {\bibinfo {volume} {67}},\ \bibinfo {pages} {181} (\bibinfo {year}
  {1991})}\BibitemShut {NoStop}%
\bibitem [{\citenamefont {Parker}\ \emph {et~al.}(2018)\citenamefont {Parker},
  \citenamefont {Yu}, \citenamefont {Zhong}, \citenamefont {Estey},\ and\
  \citenamefont {M{\"u}ller}}]{parker2018measurement}%
  \BibitemOpen
  \bibfield  {author} {\bibinfo {author} {\bibfnamefont {R.~H.}\ \bibnamefont
  {Parker}}, \bibinfo {author} {\bibfnamefont {C.}~\bibnamefont {Yu}}, \bibinfo
  {author} {\bibfnamefont {W.}~\bibnamefont {Zhong}}, \bibinfo {author}
  {\bibfnamefont {B.}~\bibnamefont {Estey}},\ and\ \bibinfo {author}
  {\bibfnamefont {H.}~\bibnamefont {M{\"u}ller}},\ }\bibfield  {title}
  {\bibinfo {title} {Measurement of the fine-structure constant as a test of
  the standard model},\ }\href {https://doi.org/10.1126/science.aap7706}
  {\bibfield  {journal} {\bibinfo  {journal} {Science}\ }\textbf {\bibinfo
  {volume} {360}},\ \bibinfo {pages} {191} (\bibinfo {year}
  {2018})}\BibitemShut {NoStop}%
\bibitem [{\citenamefont {Morel}\ \emph {et~al.}(2020)\citenamefont {Morel},
  \citenamefont {Yao}, \citenamefont {Clad{\'e}},\ and\ \citenamefont
  {Guellati-Kh{\'e}lifa}}]{morel2020determination}%
  \BibitemOpen
  \bibfield  {author} {\bibinfo {author} {\bibfnamefont {L.}~\bibnamefont
  {Morel}}, \bibinfo {author} {\bibfnamefont {Z.}~\bibnamefont {Yao}}, \bibinfo
  {author} {\bibfnamefont {P.}~\bibnamefont {Clad{\'e}}},\ and\ \bibinfo
  {author} {\bibfnamefont {S.}~\bibnamefont {Guellati-Kh{\'e}lifa}},\
  }\bibfield  {title} {\bibinfo {title} {Determination of the fine-structure
  constant with an accuracy of 81 parts per trillion},\ }\href
  {https://doi.org/10.1038/s41586-020-2964-7} {\bibfield  {journal} {\bibinfo
  {journal} {Nature}\ }\textbf {\bibinfo {volume} {588}},\ \bibinfo {pages}
  {61} (\bibinfo {year} {2020})}\BibitemShut {NoStop}%
\bibitem [{\citenamefont {M{\'e}noret}\ \emph {et~al.}(2018)\citenamefont
  {M{\'e}noret}, \citenamefont {Vermeulen}, \citenamefont {Le~Moigne},
  \citenamefont {Bonvalot}, \citenamefont {Bouyer}, \citenamefont {Landragin},\
  and\ \citenamefont {Desruelle}}]{menoret2018gravity}%
  \BibitemOpen
  \bibfield  {author} {\bibinfo {author} {\bibfnamefont {V.}~\bibnamefont
  {M{\'e}noret}}, \bibinfo {author} {\bibfnamefont {P.}~\bibnamefont
  {Vermeulen}}, \bibinfo {author} {\bibfnamefont {N.}~\bibnamefont
  {Le~Moigne}}, \bibinfo {author} {\bibfnamefont {S.}~\bibnamefont {Bonvalot}},
  \bibinfo {author} {\bibfnamefont {P.}~\bibnamefont {Bouyer}}, \bibinfo
  {author} {\bibfnamefont {A.}~\bibnamefont {Landragin}},\ and\ \bibinfo
  {author} {\bibfnamefont {B.}~\bibnamefont {Desruelle}},\ }\bibfield  {title}
  {\bibinfo {title} {Gravity measurements below 10- 9 g with a transportable
  absolute quantum gravimeter},\ }\href
  {https://doi.org/10.1038/s41598-018-30608-1} {\bibfield  {journal} {\bibinfo
  {journal} {Scientific reports}\ }\textbf {\bibinfo {volume} {8}},\ \bibinfo
  {pages} {12300} (\bibinfo {year} {2018})}\BibitemShut {NoStop}%
\bibitem [{\citenamefont {Rosi}\ \emph {et~al.}(2014)\citenamefont {Rosi},
  \citenamefont {Sorrentino}, \citenamefont {Cacciapuoti}, \citenamefont
  {Prevedelli},\ and\ \citenamefont {Tino}}]{rosi2014precision}%
  \BibitemOpen
  \bibfield  {author} {\bibinfo {author} {\bibfnamefont {G.}~\bibnamefont
  {Rosi}}, \bibinfo {author} {\bibfnamefont {F.}~\bibnamefont {Sorrentino}},
  \bibinfo {author} {\bibfnamefont {L.}~\bibnamefont {Cacciapuoti}}, \bibinfo
  {author} {\bibfnamefont {M.}~\bibnamefont {Prevedelli}},\ and\ \bibinfo
  {author} {\bibfnamefont {G.}~\bibnamefont {Tino}},\ }\bibfield  {title}
  {\bibinfo {title} {Precision measurement of the newtonian gravitational
  constant using cold atoms},\ }\href {https://doi.org/10.1038/nature13433}
  {\bibfield  {journal} {\bibinfo  {journal} {Nature}\ }\textbf {\bibinfo
  {volume} {510}},\ \bibinfo {pages} {518} (\bibinfo {year}
  {2014})}\BibitemShut {NoStop}%
\bibitem [{\citenamefont {Asenbaum}\ \emph {et~al.}(2020)\citenamefont
  {Asenbaum}, \citenamefont {Overstreet}, \citenamefont {Kim}, \citenamefont
  {Curti},\ and\ \citenamefont {Kasevich}}]{asenbaum2020atom}%
  \BibitemOpen
  \bibfield  {author} {\bibinfo {author} {\bibfnamefont {P.}~\bibnamefont
  {Asenbaum}}, \bibinfo {author} {\bibfnamefont {C.}~\bibnamefont
  {Overstreet}}, \bibinfo {author} {\bibfnamefont {M.}~\bibnamefont {Kim}},
  \bibinfo {author} {\bibfnamefont {J.}~\bibnamefont {Curti}},\ and\ \bibinfo
  {author} {\bibfnamefont {M.~A.}\ \bibnamefont {Kasevich}},\ }\bibfield
  {title} {\bibinfo {title} {Atom-interferometric test of the equivalence
  principle at the 10- 12 level},\ }\href
  {https://doi.org/10.1103/PhysRevLett.125.191101} {\bibfield  {journal}
  {\bibinfo  {journal} {Physical Review Letters}\ }\textbf {\bibinfo {volume}
  {125}},\ \bibinfo {pages} {191101} (\bibinfo {year} {2020})}\BibitemShut
  {NoStop}%
\bibitem [{\citenamefont {Du}\ \emph {et~al.}(2017)\citenamefont {Du},
  \citenamefont {Yue}, \citenamefont {Liang}, \citenamefont {Li}, \citenamefont
  {Yan},\ and\ \citenamefont {Zhu}}]{PhysRevA.95.043608}%
  \BibitemOpen
  \bibfield  {author} {\bibinfo {author} {\bibfnamefont {Y.-X.}\ \bibnamefont
  {Du}}, \bibinfo {author} {\bibfnamefont {X.-X.}\ \bibnamefont {Yue}},
  \bibinfo {author} {\bibfnamefont {Z.-T.}\ \bibnamefont {Liang}}, \bibinfo
  {author} {\bibfnamefont {J.-Z.}\ \bibnamefont {Li}}, \bibinfo {author}
  {\bibfnamefont {H.}~\bibnamefont {Yan}},\ and\ \bibinfo {author}
  {\bibfnamefont {S.-L.}\ \bibnamefont {Zhu}},\ }\bibfield  {title} {\bibinfo
  {title} {Geometric atom interferometry with shortcuts to adiabaticity},\
  }\href {https://doi.org/10.1103/PhysRevA.95.043608} {\bibfield  {journal}
  {\bibinfo  {journal} {Phys. Rev. A}\ }\textbf {\bibinfo {volume} {95}},\
  \bibinfo {pages} {043608} (\bibinfo {year} {2017})}\BibitemShut {NoStop}%
\bibitem [{\citenamefont {Kotru}\ \emph {et~al.}(2015)\citenamefont {Kotru},
  \citenamefont {Butts}, \citenamefont {Kinast},\ and\ \citenamefont
  {Stoner}}]{PhysRevLett.115.103001}%
  \BibitemOpen
  \bibfield  {author} {\bibinfo {author} {\bibfnamefont {K.}~\bibnamefont
  {Kotru}}, \bibinfo {author} {\bibfnamefont {D.~L.}\ \bibnamefont {Butts}},
  \bibinfo {author} {\bibfnamefont {J.~M.}\ \bibnamefont {Kinast}},\ and\
  \bibinfo {author} {\bibfnamefont {R.~E.}\ \bibnamefont {Stoner}},\ }\bibfield
   {title} {\bibinfo {title} {Large-area atom interferometry with
  frequency-swept raman adiabatic passage},\ }\href
  {https://doi.org/10.1103/PhysRevLett.115.103001} {\bibfield  {journal}
  {\bibinfo  {journal} {Phys. Rev. Lett.}\ }\textbf {\bibinfo {volume} {115}},\
  \bibinfo {pages} {103001} (\bibinfo {year} {2015})}\BibitemShut {NoStop}%
\bibitem [{\citenamefont {Madasu}(2023)}]{Madasu2023Thesis}%
  \BibitemOpen
  \bibfield  {author} {\bibinfo {author} {\bibfnamefont {C.~S.}\ \bibnamefont
  {Madasu}},\ }\emph {\bibinfo {title} {Experimental investigation of
  non-Abelian artificial gauge fields: from {SU(2)} to {SU(3)}}},\ \href
  {https://doi.org/10.32657/10356/165915} {Ph.D. thesis},\ \bibinfo  {school}
  {Nanyang Technological University} (\bibinfo {year} {2023})\BibitemShut
  {NoStop}%
\bibitem [{\citenamefont {Leroux}(2017)}]{Frederic2017Thesis}%
  \BibitemOpen
  \bibfield  {author} {\bibinfo {author} {\bibfnamefont {F.~J. F.~M.}\
  \bibnamefont {Leroux}},\ }\emph {\bibinfo {title} {Non-Abelian Geometrical
  Quantum Gate Operation in an Ultracold Strontium Gas}},\ \href
  {https://scholarbank.nus.edu.sg/handle/10635/137191} {Ph.D. thesis},\
  \bibinfo  {school} {National University of Singapore} (\bibinfo {year}
  {2017})\BibitemShut {NoStop}%
\end{thebibliography}%

\title{Supplemental Material for "Geometric Ramsey Interferometry with a Tripod Scheme"}
\maketitle

\section{Supplemental Material for "Geometric Ramsey Interferometry with a Tripod Scheme"}
\subsection{Sample Preparation}

We prepare an ultracold gas of $^{87}\textrm{Sr}$ atoms using laser cooling followed by evaporative cooling in a crossed beam optical-dipole trap. The two stage magneto-optical trap, operating on $^1S_0\to\,^{1}P_1$ dipole-allowed transition at $461\,$nm followed by the $^1S_0 \to\,^3 P_1$ intercombination line at $689\,$nm, cools the atomic sample of around $20\times 10^6$ atoms to a temperature of $3.3(3) \mu K$, see Refs. \cite{chaneliere2008three, Tao2015HighFlux} for details. After the laser cooling stage, about $2.5 \times 10^6$ atoms are loaded into a crossed-beams optical-dipole trap. An optical pumping sequence pumps the atoms in positive $m_F$ magnetic sub-states of the ground state to the $m_F=9/2$ stretched state, whereas the negative $m_F$ sub-states are left untouched to aid the subsequent evaporative cooling stage. Forced evaporative cooling is performed for $5.5\,$s by exponentially lowering the powers of the dipole trap beams \cite{hasan2022anisotropic}. We obtain a Fermi gas comprised of $N=4.5(2)\times10^4$ atoms in the $m_F=9/2$ stretched Zeeman sublevel at a temperature of $T_0= 50(3)\,$nK. After evaporative cooling, the optical dipole trap is switched off, and a magnetic field bias of 67 G is turned on to isolate a tripod scheme on the $^1S_0, F_g=9/2 \to\,^3 P_1, F_e=9/2$ hyperfine manifold of the intercombination line at $689\,$nm \cite{Hasan_2022}. 

\subsection{Gaussian pulses and population in the dressed-state basis}

\begin{figure}
    \centering
    \includegraphics[scale=1]{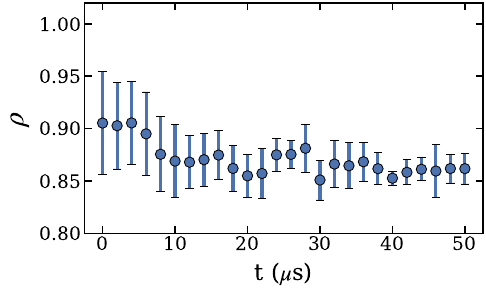}
    \caption{Fraction of total number of atoms $\rho$ in the three Gaussian distributions as a function of time during the pulse sequence}
    \label{fig:enter-label}
\end{figure}

The Gaussian pulses are produced by modulating the RF power driving the acousto-optic modulators of the tripod beams using arbitrary waveform generators \cite{Madasu2023Thesis}. The pulse sequence and the initial state preparation are such that the atoms remain in the dark-state subspace if the adiabatic condition is fulfilled. Moreover, under the adiabatic conditions, the accessible momenta of an atom are limited to $p$, $p+2\hbar k\hat{y}$, and $p+\hbar k(\hat{y}-\hat{x})$ associated respectively to the internal states $\ket{m_F=9/2}\equiv\ket{3}$, $\ket{m_F=5/2}\equiv\ket{1}$, and $\ket{m_F=7/2}\equiv\ket{2}$. Here, $p$ is the initial momentum, its distribution is given by the temperature of the gas. Hence, after the time of flight, one expects the atoms to be located on three Gaussian distributions centered on momenta $0$, $2\hbar k\hat{y}$, and $\hbar k(\hat{y}-\hat{x})$, see for example Fig. 2a-c in the main text. If unwanted diabatic transition occurs, the bright states are populated leading to spontaneous emission and loss of coherence. In this case, the accessible momenta are not limited to the three former Gaussian distributions but spread over areas of $\hbar k$ width. As a result, the populations of the three Gaussian pulses are depleted. In Fig. \ref{fig:enter-label}, we plot the fraction of the three Gaussian distribution $\rho$ as a function of the time during the pulse sequence for the experimental data depicted in Fig. 2d in the main text. We note that for the initial state preparation, we find $\rho=0.91(5)$ instead of $1$, indicating an imperfect background cancellation, and consequently a bias in the value of $\mu=\langle \rho \rangle$. Taking this bias into account we find that on average $97(2)\%$(\textit{i.e.,} $\mu = 0.97(2)$) of the atoms are in the dark states subspace throughout the Ramsey interferometer. However, as $\rho$ is weakly decreasing, we can conclude that a moderate diabatic coupling occurs during the pulse sequences. 

\subsection{Detection and Imaging}

After the tripod experimental pulses, the ground state population is extracted by exploiting the momentum-spin coupling between the ground states. After a time-of flight of TOF = 9 ms, the velocity distribution of the ultracold gas is imaged using a fluorescence imaging system. We switch on an intense retro-reflected resonant 461 nm laser beam for $20\,\mu$s. The beam strongly saturates the atomic transition (saturation parameter $s = 15$), producing a fluorescence signal that weakly depends on the probe power and density of our sample. The fluorescence signal is collected on a EMCCD camera through a 2.5 magnification imaging system normal to the plane of the tripod beams. The spatial resolution of the imaging system is 13 $\mu$m, ultimately limited by the 16 $\mu$m camera pixel size. The velocity resolution of our TOF images is 0.1$v_r$. A signal-to-noise ratio of one on a camera pixel corresponds to a fluorescence signal given by $\approx$ 70 atoms, see Ref. \cite{Madasu2023Thesis} for more details.

In the final sample, the atoms in the ground state are distributed among the Zeeman sublevels $m_F = 9/2$ and in the $m_F < 0$. The atoms in $m_F < 0$ are unaffected by the tripod beams and thus remain spectators introducing a bias in the population measurement of $m_F = 9/2$ at $p=0$ as the fluorescence images are sensitive to the velocity distribution but not to the value of $m_F$. To remove this bias, we measure the fraction of the total number of atoms in $m_F < 0$ states by performing a STIRAP that transfers the atoms from $m_F = 9/2$ to $m_F = 5/2$ with almost perfect efficiency imparting a velocity of $2v_r$. We found that 51(4)\% of the total number of atoms are in $mF < 0$ states. We disregard this fraction of the total number of atoms in all the population measurements.\\

\subsection{Models and numerical simulations}
\begin{figure}
    \centering
    \includegraphics{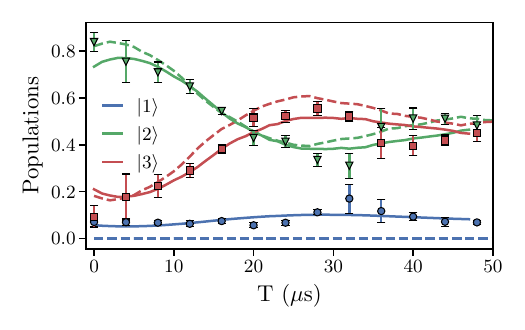}
    \caption{Comparison of experimental data with thermal averaged optical-Bloch simulation (plain curves) and gauge field approach (dashed curves).}
    \label{supp_figure2}
\end{figure}
The numerical simulations presented in the figures in the main text are obtained using a model valid in the adiabatic limit where the evolution is constrained within the dark state subspace. The Hamiltonian reads 
\begin{equation}
    \hat{H}=\frac{\hat{\textbf{p}}^2\otimes\mathds{1}}{2m}-\frac{\hat{\textbf{A}}\cdot\hat{\textbf{p}}}{m}+\hat{Q}+\hat{w},
\end{equation}
as Eq. (2) in the main text. The thermal averaging is performed in a semi-classical limit, where the external degree of freedom are treated as classical quantities. The unitary operator becomes
\begin{equation}
	\hat{U}(t)=\mathcal{T}\exp{\left[-i\int_0^{t}\left(\hat{Q}(t^\prime)+\hat{w}(t^\prime)-v_x A_x-v_y A_y\right)\textrm{d}t^\prime\right]},
	\label{unitaryTSupp}
\end{equation}
where $(v_x,v_y)$ are the atomic velocity components in the plane of the atomic beams. The numerical simulations are performed using 2000 velocities randomly extracted from a Maxwell-Boltzmann distribution with a variance in agreement with the experimental temperature of $T_0=50$ nK. For each velocity component, we extract the populations of bare states from the dark states as presented in Ref. \cite{Frederic2018Non-Abelian}. Finally, we average the bare-state populations over the 2000 runs.

A second approach consists of solving optical-Bloch equations (OBEs) in the atomic bare-state basis. Here, the interaction Hamiltonian, in $\{\ket{e}, \ket{1}, \ket{2}, \ket{3}\}$ representation, reads,
\begin{equation}
    \hat{H_I} (t) = \frac{\hbar}{2} \left(\begin{array}{cccc}
 		0 & \Omega_1(t) & \Omega_2(t) & \Omega_3(t)\\
 		\Omega_1^*(t)& -2\delta_1 & 0 & 0\\
 		\Omega_2^*(t)& 0 & -2\delta_2 & 0\\
 		\Omega_3^*(t)& 0 & 0 & -2\delta_3
 	\end{array}\right)
    \label{internalHam}
\end{equation}
where $\Omega_i(t)$ and $\delta_i$ for $i=\{1,2,3\}$ are time-dependent Rabi frequencies and detunings of the tripod transitions. To compute the OBEs, we add the necessary relaxation of the excited state population and coherence \cite{Frederic2017Thesis}. The resolution of the OBEs is done in the atom rest frame, where the detunings are given by \cite{hasan2022anisotropic}
\begin{align}
    \label{detunings_thermal}
    \delta_1 &= 4\omega_r+k v_y\nonumber\\
    \delta_2 &= 2\omega_r-k v_x\\
    \delta_3 &= -k v_y.\nonumber
\end{align}
The terms, proportional to the recoil frequency in the right-hand side of Eqs. (\ref{detunings_thermal}), are due to photon redistributions among the tripod beams, considered here as plane waves. The terms, proportional to the atomic velocity components in the right-hand side of Eqs. (\ref{detunings_thermal}), take into account the Doppler shifts, for an atom moving at a velocity $\boldsymbol{v} = v_x\hat{x}+v_y\hat{y}$. 
As in the adiabatic model, we solve OBEs using 2000 velocities derived randomly from Maxwell-Boltzmann distribution corresponding to the temperature $T_0=50$ nK of the gas. 

For the slow pulses, \textit{i.e.} $\sigma_t=8\mu$s and above, the adiabatic and OBE models match well with each other. For the parameters used in the experiment, \textit{i.e.} $\sigma_t=2.5\mu$s, the two models give slightly different results as shown in Fig. \ref{supp_figure2}. Here the dashed (plain) curves correspond to the adiabatic (OBEs) model. As expected, the OBEs model agrees better with the experiment. In particular, the non-zero population of the $\ket{2}$ state is well captured with the OBEs model, thus corresponding to diabatic contribution. 


\end{document}